\documentclass{article}	

\usepackage{amsmath,amsthm,amssymb,cite,enumerate} 

\usepackage[dvips]{color}
\usepackage{authblk} 

\def \BEA { \begin{eqnarray}}
\def \EEA {\end{eqnarray}}
\def \BE {\begin{equation}}
\def \EE {\end{equation}}
\def\d{\mathrm{d}}

\def \WDS #1 {\mbox{$\Phi_{#1}^{S}$}}
\def \WDA #1 {\mbox{$\Phi_{#1}^{A}$}}
\def \WD #1 {\mbox{$\Phi_{#1}$}}

\newcommand{\OO}[1] {{O}(r^{-#1})}
\newcommand{\OOp}[1] {{O}(r^{#1})}

\newcommand{\oo}[1] {o(r^{-#1})}

\def\a{\alpha}
\def\b{\beta}

\def \mi {\stackrel{i}{m}}
\def \mj {\stackrel{j}{m}}
\def \mk {\stackrel{k}{m}}
\def \mr {\stackrel{r}{m}}
\def \ms {\stackrel{s}{m}}
\def \mz {\stackrel{z}{m}}
\def \mq {\stackrel{q}{m}}
\def \mo {\stackrel{o}{m}}
\def \mD {\stackrel{2}{m}}
\def \mT {\stackrel{3}{m}}
\def \mC {\stackrel{4}{m}}

\def \mio #1 {\mi_{#1}\ ^{  \! \! \! \! 0}} 
\def \mjo #1 {\mj_{#1}\ ^{  \! \! \! \! 0}} 
\def \mko #1 {\mk_{#1}\ ^{  \! \! \! \! 0}} 
\def \mro #1 {\mr_{#1}\ ^{  \! \! \! \! 0}} 
\def \mso #1 {\ms_{#1}\ ^{  \! \! \! \! 0}} 
\def \mpo #1 {\mp_{#1}\ ^{  \! \! \! \! 0}} 
\def \mzo #1 {\mz_{#1}\ ^{  \! \! \! \! 0}} 
\def \mqo #1 {\mq_{#1}\ ^{  \! \! \! \! 0}} 
\def \moo #1 {\mo_{#1}\ ^{  \! \! \! \! 0}} 
\def \mDo #1 {\mD_{#1}\ ^{  \! \! \! \! 0}} 
\def \mTo #1 {\mT_{#1}\ ^{  \! \! \! \! 0}} 
\def \mCo #1 {\mC_{#1}\ ^{  \! \! \! \! 0}} 

\def \miJ #1 {\mi_{#1}\ ^{  \! \! \! \! (1)}} 
\def \mjJ #1 {\mj_{#1}\ ^{  \! \! \! \! (1)}} 
\def \mkJ #1 {\mk_{#1}\ ^{  \! \! \! \! (1)}} 
\def \mrJ #1 {\mr_{#1}\ ^{  \! \! \! \! (1)}}

\def \bl {\mbox{\boldmath{$\ell$}}}

\def \bn {\mbox{\boldmath{$n$}}}

\def \hbm #1 {\mbox{\boldmath{$\hat m^{(#1)}$}}}
\def \bm {\mbox{\boldmath{$m$}}}

\def \hx {\mbox{\boldmath{$\hat\xi$}}}
\def \he {\mbox{\boldmath{$\eta$}}}

\def \Ms {\stackrel{s}{M}}

\newcommand{\be}{\begin{equation}}
\newcommand{\ee}{\end{equation}}
\newcommand{\beqn}{\begin{eqnarray}}
\newcommand{\eeqn}{\end{eqnarray}}
\newcommand{\pa}{\partial}
\newcommand{\ba}{\begin{array}}
\newcommand{\ea}{\end{array}}

\newcommand{\e}{\epsilon}

\def \nb {\beta}
\def \ng {\gamma}
\def \nd {\delta}

\def \BEAH {\begin{eqnarray*}}
\def \EEAH {\end{eqnarray*}}
\def \BEA {\begin{eqnarray}}
\def \EEA {\end{eqnarray}}
\def \BDM {\begin{displaymath}}
\def \EDM {\end{displaymath}}

\def \Ms {\stackrel{s}{M}}

\def \T {\bigtriangleup  }

\newcommand{\M}[3] {{\stackrel{#1}{M}}_{{#2}{#3}}}
\newcommand{\m}[3] {\!{\stackrel{\hspace{.3cm}#1}{m}}_{\!{#2}{#3}}\,}

\begin{document}

\title{On the uniqueness of the Myers-Perry spacetime as a type II(D) solution in six dimensions}


\author[1,2]{Marcello Ortaggio\thanks{ortaggio(at)math(dot)cas(dot)cz}}
\affil[1]{Institute of Mathematics of the Czech Academy of Sciences, \newline \v Zitn\' a 25, 115 67 Prague 1, Czech Republic}
\affil[2]{Instituto de Ciencias F\'{\i}sicas y Matem\'aticas, Universidad Austral de Chile, \newline Edificio Emilio Pugin, cuarto piso, Campus Isla Teja, Valdivia, Chile}

\date{\today}

\maketitle

\begin{abstract}

We study the class of vacuum (Ricci flat) six-dimensional spacetimes admitting a non-degenerate multiple Weyl aligned null direction $\bl$, thus being of Weyl type II or more special. Subject to an additional assumption on the asymptotic fall-off of the Weyl tensor, we prove that these spacetimes can be 
completely classified in terms of the two eigenvalues of the (asymptotic) twist matrix of  $\bl$ and of a discrete parameter $U^0=\pm 1/2, 0$. All solutions turn out to be Kerr-Schild spacetimes of type D and reduce to a family of ``generalized'' Myers-Perry metrics (which include limits and analytic continuations of the original Myers-Perry black hole metric, such as certain NUT spacetimes). A special subcase corresponds to twisting solutions with zero shear. In passing, limits connecting various branches of solutions are briefly discussed.

\end{abstract}



\tableofcontents

\section{Introduction}

\subsection{Background}

In the context of exact solutions to Einstein's equations, the study of algebraically special spacetimes has been a fruitful line of research for many years. Most notably, it led to the discovery of the Kerr metric \cite{Kerr63}. In the vacuum case, the Goldberg-Sachs \cite{GolSac62} theorem is a milestone for the integration of the Newman-Penrose equations \cite{NP}, and enables one to classify algebraically special solutions as either diverging ($\rho\neq0$) or non-diverging ($\rho=0$, i.e., Kundt). 
In both cases, all algebraically special types (i.e., type II or more special) can occur (cf. \cite{Stephanibook} and references therein). The Kerr black hole and its generalizations belong to the subfamily of diverging type D vacua. Remarkably, all type D vacua can be fully classified and integrated, and the general metric contains only {\em constants} of integration \cite{Kinnersley69}.

An extension of the Petrov classification to higher dimensions has been put forward in \cite{Coleyetal04} (cf. \cite{OrtPraPra13rev} for a review). In arbitrary dimensions, the type II condition of \cite{Coleyetal04} can be expressed  as \cite{Ortaggio09}
\be
	\ell_{[e}C_{a]b[cd}\ell_{f]}\ell^b=0 ,
	\label{mWAND}
\ee
where $\bl$ is a null vector field. The null direction defined by $\bl$ is a {\em multiple Weyl aligned null direction} (mWAND). Generalizations of the Schwarzschild and Kerr solutions to arbitrary dimensions have been known for some time \cite{Tangherlini63,MyePer86}, and they are of type D \cite{Pravdaetal04,Hamamotoetal07,PraPraOrt07} (i.e., they admit {\em two} distinct mWANDs). One might hope that, similarly as in four dimensions, all type D vacua can be found in any dimensions. More generally, it appears of interest to study and classify all vacuum solutions of type II or more special, i.e., those satisfying \eqref{mWAND}.

In five dimensions, a first step in this direction was an extension of the Goldberg-Sachs theorem \cite{Pravdaetal04,DurRea09,Ortaggioetal12}.\footnote{Only one direction of the Goldberg-Sachs theorem has been investigated systematically in higher dimensions, that is, what the existence of a mWAND implies in a vacuum spacetime. This is the only part of the theorem relevant to this paper, and hereafter this will be understood. It should also be noted that a different formulation of the higher-dimensional Goldberg-Sachs theorem has been studied in \cite{Taghavi-Chabert11,Taghavi-Chabert11b}.}  This defines three branches of non-Kundt solutions \cite{Ortaggioetal12}, according to the possible rank (3, 2, or 1) of the optical matrix (defined below in \eqref{L_def}). All such solutions (including a cosmological constant) have been fully classified in \cite{deFGodRea15,Wylleman15,deFGodRea16} (see also \cite{ParWyl11,ReaGraTur13} for earlier results in special cases). In particular, in the {\em full-rank} case \cite{deFGodRea15}, they essentially reduce the Myers-Perry black hole solution with a cosmological constant \cite{HawHunTay99} and certain limits thereof, thus being of type D and specified just by a few parameters. This is in contrast with the existence of diverging solutions of type II, III and N in four dimensions.

In more than five dimensions, a complete extension of the Goldberg-Sachs theorem has not yet been achieved (see, however, \cite{Pravdaetal04,OrtPraPra09,DurRea09,OrtPraPra09b,OrtPraPra13}). Moreover, there exist a qualitative difference between $n=4,5$ and any higher dimensions in the structure of the Weyl tensor, amounting to new ``degrees of freedom'' in the purely spatial components $C_{ijkl}$ \cite{PraPraOrt07,OrtPra14}.\footnote{As a consequence, for example, for $n\ge 6$ there exist static vacuum black holes with horizons of non-constant curvature \cite{Birmingham99}.} This means that the integration of the Newman-Penrose equations is in general more complicated, and may require new techniques. However, earlier results in various special cases \cite{Pravdaetal04,PodOrt06,OrtPraPra09b,OrtPraPra13} indicate that a ``uniqueness'' result similar to that of \cite{deFGodRea15} may hold also in more than five dimensions dimensions, when the optical matrix has full-rank. This is what we will prove in this paper in six dimensions, under an additional assumption explained below (eq.~\eqref{assump_C2}). We believe that, together with \cite{deFGodRea15}, the methods and the results of this paper will give enough insight to tackle the case of general dimensions, at least in the full-rank case. Further motivation for studying six dimensional spacetimes is that only {\em even} dimensions allow for shearfree WANDs with non-zero twist \cite{OrtPraPra07}. Since the Myers-Perry metric is shearing for $n\ge5$ \cite{MyePer86,FroSto03,PraPraOrt07,OrtPraPra09}, this might thus lead to new solutions not present in five dimensions. We will find that such solutions indeed exist, but they turn out not to be new (section~\ref{subsubsec_shearfree_twisting}).

In the remaining part of this section we describe the assumptions made in the present paper and we summarize our results, which will be proved in the rest of the paper. In section~\ref{sec_r_dep}, the $r$-dependence of the non-zero Newman-Penrose quantities is obtained (in some cases only asymptotically, which suffices for our purposes), together with the ``transverse'' equations, to be employed subsequently (the results of sections~\ref{sec_r_dep} and \ref{subsec_coord_u} are given in arbitrary dimension $n>4$, since no significant simplification would be achieved by fixing a particular value of $n$). In section~\ref{sec_adapted} we define adapted coordinates (not completely specified yet) and a natural parallely transported frame. The complete integration of the field equations is carried out in sections~\ref{sec_generic} and \ref{sec_shearfree} for the two possible cases $b_{23}^2\neq b_{45}^2$ and $b_{23}^2=b_{45}^2$, respectively. Appendix~\ref{app_MP} describes a family of ``generalized'' Myers-Perry metrics (giving various coordinate systems and discussing limits and analytic continuations connecting various branches of the solutions). Appendix~\ref{app_NP} summarizes the higher dimensional Newman-Penrose formalism used throughout the paper.

\subsection{Assumptions}

\label{subsec_assump}

Let us consider a vacuum (Ricci flat) $n$-dimensional spacetime of type II or more special ($n>4$). In a spacetime admitting a mWAND, there always exists a {\em geodesic} mWAND \cite{DurRea09}. With no loss of generality we can thus assume $\bl$ to satisfy \eqref{mWAND}  and to be geodesic and affinely parametrized, i.e., $\ell_{a;b}\ell^b=0$. This implies (cf. section~2.2 of \cite{OrtPraPra07}) that the rank the {\em optical matrix}
\be
  L_{ij}=\ell_{a;b}m_{(i)}^am_{(j)}^b ,
	\label{L_def}
\ee
is a frame-independent property, for any choice of a frame adapted to $\bl$ (the $n$ frame vectors $\bm_{(a)}$ consists of two null vectors $\bl\equiv\bm_{(0)}$,  $\bn\equiv\bm_{(1)}$ and $n-2$ orthonormal spacelike vectors $\bm_{(i)} $, with $a, b\ldots=0,\ldots,n-1$ while $i, j  \ldots=2,\ldots,n-1$ \cite{Coleyetal04,OrtPraPra13rev}).

In this paper we focus on the {\em non-degenerate}  (i.e., full-rank) case, namely we assume hereafter
\be
 \det L\neq 0 .
 \label{detL}
\ee
Thanks to \cite{Pravdaetal04,Kubicek_thesis}, this implies that the Weyl type can only be II or D (or O, in the trivial case of Minkowski's space).

For convenience, throughout the paper we take our frame to be parallelly propagated along $\bl$ \cite{OrtPraPra07} (cf.~\eqref{parall_transp}), and we define an affine parameter $r$ such that
\be
 \bl=\pa_r .
 \label{l}
\ee

Under the above assumptions, the higher-dimensional Sachs equation $DL=-L^2$ \cite{Pravdaetal04,OrtPraPra07} fixes the $r$-dependence of $L$ as \cite{OrtPraPra09,OrtPraPra09b} (cf.~\cite{NP} in 4D)
\be
 L^{-1}=r{I}-b , 
 \label{inverse}
\ee
where $I$ is the identity matrix, and the ``integration matrix'' $b$ satisfies $Db=0$. Without loss of generality, we can restrict ourselves to the case $b\neq0$, for the case $b=0$ reduces to the Robinson-Trautman metrics, already studied in \cite{PodOrt06}.

As an additional assumption, we further require that the fall-off behaviour of the spatial part of the Weyl tensor for large~$r$ be ``fast enough'', namely 
\be
  C_{ijkl}=\oo{2} . 
 \label{assump_C2}
\ee
The motivation for this choice is twofold. From a physical viewpoint, such a fall-off is necessary for asymptotic flatness \cite{OrtPraPra09b,GodRea12}, and is thus a natural condition to consider.\footnote{In mathematical terms, imposing \eqref{assump_C2} means that certain integration functions (when fixing the Weyl $r$-dependence from the Bianchi identities) are set to be zero \cite{OrtPra14} (cf. also, e.g., section~5 of \cite{PraPra08} and section~4 of \cite{OrtPraPra13}). A price to pay for this simplification is that some solutions admitting a (geodesic, non-degenerate) mWAND, such as static black holes with a generic Einstein horizon \cite{Birmingham99} or rotating black holes with non-zero NUT \cite{CheLuPop06}, will be excluded from the spacetimes under consideration (they both violate \eqref{assump_C2}, as follows from \cite{PodOrt06} and \cite{Hamamotoetal07}, respectively).} Additionally, it allows for a partial extension of the Goldberg-Sachs theorem to higher dimensions \cite{OrtPraPra09b}, which results in a significant simplification of the Newman-Penrose equations one has to solve. Namely, from the Bianchi identities it follows \cite{OrtPraPra09b} that $b_{(ij)}\propto\delta_{ij}$, which is equivalent (when \eqref{detL} holds) to the so called ``optical constraint'' \cite{OrtPraPra09,OrtPraPra13}. By a shift of $r$ one can thus set $b_{(ij)}=0$, so that, without loss of generality, from now on in \eqref{inverse} we take
\be
	b_{ij}=b_{[ij]}\neq 0 .
\ee	
We observe that $b_{[ij]}$ gives the twist matrix of $\bl$ at the leading order in $1/r$ (as follows from \eqref{inverse}).

\subsection{Summary of results}

\label{subsec_summary}

We will prove that, in six dimensions, all vacuum spacetimes admitting a geodesic mWAND and further satisfying \eqref{detL} and \eqref{assump_C2} can be fully classified in terms of (the moduli of) the two eigenvalues $b_{23}$ and $b_{45}$ of the asymptotic twist matrix $b_{[ij]}$ (cf.~\eqref{b_eigenf}) and of a discrete parameter $U^0 =\pm1/2, 0$ (cf.~\eqref{frame_coefficients}, \eqref{U})  as follows.  

\begin{enumerate}
	\item If $\d b_{23}\neq0\neq\d b_{45}$, for $U^0\neq0$ the spacetime is equivalent to the doubly spinning Myers-Perry metric with {\em unequal} rotation parameters \eqref{MP_6D_CLP}.  In the special subcase $U^0=0$, one obtains instead the metric~\eqref{generic_U0=0}.

	\item If $\d b_{23}=0\neq\d b_{45}$ then
		\begin{enumerate}
			\item if $b_{23}\neq0$, one has the doubly spinning Myers-Perry metric with {\em equal} rotation parameters \eqref{MP_6D_equal};
					
			\item if $b_{23}=0$, one has the Myers-Perry metric with a single spin \eqref{MP_6D_single} for  $U^0\neq0$, and its limit \eqref{strange} for $U^0=0$. 
  
		\end{enumerate}
  \item If $\d b_{23}=0=\d b_{45}$ then two cases arise. 					
			\begin{enumerate}
			\item If $b_{23}^2+b_{45}^2\neq0$, the NUT metric~\eqref{MP_e=0} (with one or two NUT parameters), first found in \cite{ManSte04}, is obtained. In the special case $b_{23}=b_{45}$ the NUT parameters coincide and $\bl$ is {\em shearfree}.
			
			\item If $b_{23}=0=b_{45}$, one is left with the generalized Schwarzschild-Tangherlini metric~\eqref{RT} with spherical, plane or hyperbolic symmetry. 
			  
		\end{enumerate}
\end{enumerate}

It is understood that such results are local. The Myers-Perry metrics referred to above are to be understood in a ``generalized'' sense, i.e., some of those are contained in \cite{MyePer86} only up to certain analytical continuations (see appendix~\ref{app_MP} for details and for some comments on the geometrical meaning of the sign of $U^0$).  Nevertheless, similarly as in \cite{deFGodRea15}, in the various cases the generalized Myers-Perry metrics can be expressed in a unified form (i.e., with no need of any analytical continuation) at the price of dressing the line-element with an additional parameter $\e$ (which we rescale to $\pm1$).

The various line-elements will be obtained in ``Eddington-like'' coordinates $(u,r,x^\alpha)$, which are natural in the Newman-Penrose formalism. However, we will present also the corresponding ``Boyer-Lindquist'' coordinates $(t,r,x'^\alpha)$, more frequently used in the black hole literature.

\section{Fixing the $r$-dependence, and transverse equation (arbitrary $n>4$)}

\label{sec_r_dep}

\subsection{Ricci rotation coefficients and derivative operators}

\label{subsec_ricci_coeff}

So far, the frame vector $\bn$ has not been specified, except for the requirement that it be parallely transported along $\bl$. While retaining the latter condition, thanks to \eqref{detL} we can perform a null rotation which sets (cf. appendix~D.2.6 of \cite{OrtPraPra09}) 
\be
 L_{i1}=0 ,
 \label{Li1}
\ee
and thus uniquely fixes the null direction defined by $\bn$. On the other hand, we still have the freedom of $r$-independent boosts in the plane of $\bl$ and $\bn$, and $r$-independent spatial rotations of the $\bm_{(i)} $ -- this will be useful in the following. 

The $r$-dependence of the non-zero Ricci rotation coefficients is fully determined (recalling \eqref{inverse}) by the Ricci identities~(11b), (11n), (11a), (11j), (11m) and (11f) of \cite{OrtPraPra07}, and reads \cite{OrtPraPra_prep} (see also appendix~D of \cite{OrtPraPra09})
\beqn
  & & L_{1i}=L_{ji}l_{1j} , \qquad \M{i}{j}{k}=L_{lk}\m{i}{j}{l} ,  \label{L1i} \\
	& & L_{11}=\frac{1}{n-2}L^{-1}_{ji}\Phi_{ij}+l_{11} , \qquad N_{ij}=L_{kj}\left(n_{ik}-\int L^{-1}_{lk}\Phi_{li}\d r\right) , \qquad \M{i}{j}{1}=-2\int\WDA{ij} \d r+\m{i}{j}{1} , \label{L11} \\
	& & N_{i1}=\int\Psi_i\d r+n_{i 1} ,  \label{Ni1}
\eeqn
where lowercase Latin letters denote (for now arbitrary) integration functions that do not depend on $r$, with $\m{i}{j}{k}=-\m{j}{i}{k}$ and $\m{i}{j}{1}=-\m{j}{i}{1}$. (To express $L_{11}$, we used the equation $(n-2)\Phi=-D(L^{-1}_{ji}\Phi_{ij})$, which follows from (B.5,\cite{Pravdaetal04}).)

Taking the affine parameter $r$ as one of the coordinates, we can write the basis vectors as 
\be
	\bl=\pa_r , \qquad \bn=U\pa_r+X^A\pa_A , \qquad \bm_{(i)} =\omega_i\pa_r+\xi^A_i\pa_A , 
	\label{frame_coefficients}
\ee
where $U$, $X^A$, $\omega_i$ and $\xi^A_i$ are spacetime functions (to be determined), $\pa_A=\pa/\pa x^A$, and the $x^A$ represent any set of ($n-1$) scalar functions such that $(r,x^A)$ is a well-behaved coordinate system. The $r$-dependence of the functions in \eqref{frame_coefficients} can be determined using the commutators \eqref{comm_DelD} and \eqref{comm_dD}, and reads \cite{OrtPraPra_prep}
\beqn 
 & & \omega_i=-L_{1i}r+L_{ji}\omega^0_j , \qquad  \xi^A_i=L_{ji}\xi^{A0}_j  , \label{omegai} \\
 & &  X^A=X^{A0} , \qquad U=-l_{11}r-\frac{1}{n-2}\int L^{-1}_{ji}\Phi_{ij}\d r+U^0 , \label{U}
\eeqn
where a superscript $^0$ denotes $r$-independent quantities.

Clearly, the $r$-dependence of~\eqref{L11}, \eqref{Ni1} and the second of \eqref{U} is, at this stage, determined only implicitly, since some Weyl components appear there -- their $r$-dependence will be given below (at least at the leading order in $1/r$). Let us also observe that the above result holds also without the assumption \eqref{assump_C2}, and can be easily extended to include a cosmological constant \cite{OrtPra14,OrtPraPra_prep}.

\subsection{Weyl components of boost weight zero}

\label{subsec_bw0}

Using now also assumption~\eqref{assump_C2}, the $r$-dependence of the Weyl b.w.~0 components has been already studied in \cite{OrtPraPra09b,OrtPra14}\footnote{In \cite{OrtPraPra09b} it was {\em assumed} that positive powers of $r$ vanish, however this assumption is not necessary \cite{OrtPra14}.}. At the leading order in $1/r$ it reads
\beqn
	& & \WDS{ij} =\frac{\Phi_0}{n-2}\left[\frac{\delta_{ij}}{r^{n-1}}+\frac{n-1}{2(n-3)}\frac{b_{kl}b_{lk}\delta_{ij}+2b_{ik}b_{kj}}{r^{n+1}}\right]+\OOp{-n-3} , \qquad  \WDA{ij} =\frac{(n-1)\Phi_0}{(n-2)(n-3)}\frac{b_{[ij]}}{r^{n}}+ \OOp{-n-2}, \label{Phi_leading} \\ 
	& & C_{ijkm}=\frac{4\Phi_0}{(n-2)(n-3)} \frac{\delta_{i[m}\delta_{k]j}}{r^{n-1}} + \OOp{-n-1} , \label{Cijkm_leading}
\eeqn
where $\Phi_0\neq0$ is an integration function independent of $r$ (for $\Phi_0=0$ all the b.w.~0 components vanish \cite{OrtPraPra09b,OrtPraPra_prep} and the spacetime is flat). For our purposes, terms of higher order are not needed. It is just important to observe that these can be determined recursively to any desired order once the leading terms in \eqref{Phi_leading}, \eqref{Cijkm_leading} are known \cite{OrtPraPra_prep}, and do {\em not} involve any integration functions other that $\Phi_0$ and $b_{ij}$. This implies (cf. \eqref{frame_coefficients} with \eqref{inverse}, \eqref{omegai} and \eqref{U}) that {\em the full spacetime metric is uniquely determined by knowing $b_{ij}$, $l_{1i}$, $l_{11}$, $\omega^0_i$, $\xi^{A0}_i$, $X^{A0}$, $U^0$ and $\Phi_0$}.

By a suitable ($r$-independent) rescaling of the affine parameter $r$ and a corresponding rescaling (boost) of $\bl$, such that $\bl=\pa_r$ and \eqref{Li1} are preserved, one can always set (cf., e.g., \cite{Kinnersley69,deFGodRea15})
\be
	\Phi_0=\mbox{const}\neq0 .
	\label{Phi_const}
\ee	
This choice simplifies considerably several expressions to be obtained in the following.

\subsection{Weyl components of negative boost weight}

In the present class of spacetimes, the Weyl components of negative b.w. fall off as $r^{1-n}$ or faster \cite{OrtPraPra09b,OrtPra14}. Using the Bianchi identities \eqref{PII-B1} and \eqref{PII-B6} (with \eqref{inverse}, \eqref{L1i}, \eqref{omegai} and \eqref{Phi_leading}), at the leading order one finds $(n-3)\Psi_{i}=-[(n-1)l_{1i}\Phi_0+\xi^{A0}_i\Phi_{0,A}]r^{1-n}+\OO{n}$ and $(n-2)\Psi_{ijk}=\delta_{k[j}\Psi_{i]}+\OO{n}$. However, from $\Psi_i=2\Psi_{ijj}$ with \eqref{Phi_const} one arrives at
\be
 l_{1i}=0 , 
 \label{l1i=0} 
\ee
so that, in fact, $\Psi_{ijk}=\OO{n}$ and $L_{1i}=0$. To proceed, it is useful to employ also the Ricci identities \eqref{11k_NP} and \eqref{11i_NP}, which give (with \eqref{Li1} and \eqref{l1i=0})
\beqn
 & & \xi^{A0}_{[j|}b_{i|k],A}=\omega^0_{[j}\delta_{k]i}+b_{il}\m{l}{[j}{k]}+b_{l[j|}\m{l}{i}{|k]} , \label{11k} \\
 & & X^{A0}b_{ij,A}=U^0\delta_{ij}-n_{ij}-l_{11}b_{ij}-b_{kj}\m{k}{i}{1}-b_{ik}\m{k}{j}{1} . \label{11i} 
\eeqn

Using \eqref{11k}, one obtains from \eqref{PII-B1} and \eqref{PII-B9} 
\be
 \Psi_i=\frac{n-1}{n-2}\frac{\Phi_0\omega^0_i}{r^n}+\OOp{-n-1}, \qquad \Psi_{ijk}=\frac{1}{n-3}\Psi_{[i}\delta_{j]k}+\OOp{-n-1} . 
 \label{Psi_i}
\ee
Comparing the latter with \eqref{PII-B6} leads to 
\be
 \xi^{A0}_{j}b_{ik,A}=2\omega^0_{[i}\delta_{k]j}+2b_{s[i}\m{s}{k]}{j} ,
 \label{xi_B}
\ee
which will be useful for later calculations.

Next, at the leading order \eqref{PII-B4} (with \eqref{inverse}, \eqref{L1i}, \eqref{L11}, \eqref{omegai}--\eqref{Phi_leading}) gives
\be
 	l_{11}=0 , \label{l11=0} 
\ee
while at the subleading order (using \eqref{11i} and $\Psi_{[ij]}=0$) one finds 
\beqn 
 & & n_{ij}=n_{(ij)}=U^0\delta_{ij} , \label{nij} \\
 & & X^{A0}b_{ij,A}=2b_{k[i}\m{k}{j]}{1} , \label{X_B} 
\eeqn
along with $\Psi_{ij}=\OO{n}$. At the next order, comparing \eqref{PII-B4} with the trace of \eqref{B13} (recalling \eqref{Phi_leading}, \eqref{Cijkm_leading}, \eqref{Psi_i}, \eqref{nij}, \eqref{X_B}), one arrives at 
\beqn
	& &  \xi^{A0}_{[i}\omega^0_{j],A}=2U^0b_{ij}+\omega^0_k\m{k}{[i}{j]} ,	\label{xi_om_a} \\
	& & \xi^{A0}_{(j}\omega^0_{i),A}+\omega^0_{s}\m{s}{(i}{j)}=0 . \label{xi_om}
\eeqn
The latter also implies that, in fact,
\be
 \Psi_{ij}=\OOp{-n-1} . 
\label{Psi_ij}
\ee

Similarly as for components of b.w.~0, also negative b.w. components \eqref{Psi_i} and \eqref{Psi_ij} can be determined to any desired order, without involving any new integration functions \cite{OrtPraPra_prep} (the explicit form of $\Psi_{ijk}$ and $\Psi_{ij}$ will not be needed in what follows).

Using the above results and \eqref{Phi_leading}, we are thus able to summarize the leading-order $r$-dependence of all the Ricci rotation coefficients \eqref{L1i}--\eqref{Ni1}
\beqn
  & & L_{1i}=0 , \qquad \M{i}{j}{k}=\frac{\m{i}{j}{k}}{r}+\OO{2} , \qquad L_{11}=\frac{1}{n-2}\frac{\Phi_0}{r^{n-2}}+\OO{n} , \label{ricci_asym1} \\
	& & N_{ij}=U^0\left(\frac{\delta_{ij}}{r}+\frac{b_{ij}}{r^2}\right)+\OO{3} , \qquad \M{i}{j}{1}=\m{i}{j}{1}+\OOp{1-n} ,  \qquad N_{i1}=n_{i 1}+\OOp{1-n} , \label{ricci_asym2}
\eeqn
and of the derivative operators \eqref{omegai}, \eqref{U} 
\be
 \omega_i=\frac{\omega^0_i}{r}+\OO{2} , \qquad  \xi^A_i=\frac{\xi^{A0}_i}{r}+\OO{2}  , \qquad X^A=X^{A0} , \qquad U=U^0+\frac{\Phi_0}{(n-2)(n-3)}\frac{1}{r^{n-3}}+\OOp{1-n} . 
 \label{der_asym}
\ee
Recall, however, that the $r$-dependence of all the metric coefficients, except for $g^{rr}=2U+\omega_i\omega_i$, is also known in closed form, thanks to \eqref{frame_coefficients} with \eqref{inverse}, \eqref{omegai} and \eqref{U}.

\subsection{Further transverse equations from commutators and Ricci identities}

The remaining commutators \eqref{comm_dDel} and \eqref{comm_dd} (with $L_{1i}=0$, as obtained above) applied on $r$ and $x^A$ give the following set of equations  (at the leading and, in the case of \eqref{X_om}, subleading order)
\beqn
 & & n_{i1}=0 ,  \label{11c_2} \\
 & & X^{A0}\omega^0_{i,A}-\xi^{A0}_iU^0_{,A}=-\m{j}{i}{1}\omega^0_j , \label{X_om}   \\
 & & \xi^{B0}_iX^{A0}_{,B}-X^{B0}\xi^{A0}_{i,B}=\m{j}{i}{1}\xi^{A0}_j , \label{X_xi} \\
 & & \xi^{B0}_{[i}\xi^{A0}_{j],B}=X^{A0}b_{ij}+\xi^{A0}_{k}\m{k}{[i}{j]} .  \label{xi_xi} 
\eeqn

Using \eqref{nij} and \eqref{11c_2}, the Ricci identities \eqref{11h_NP}--\eqref{11p_NP} give, respectively, 
\beqn
 & & X^{A0}U^0_{,A}=0 , \label{X_U} \\
 & & \xi^{A0}_{i}U^0_{,A}=0 , \label{xi_U} \\
 & & X^{A0}\m{i}{j}{k,A}-\xi^{A0}_{k}\m{i}{j}{1,A}=-2\m{s}{[i|}{1}\m{s}{|j]}{k}-\m{i}{j}{s}\m{s}{k}{1} , \label{11o_2} \\
 & & \xi^{A0}_{[k|}\m{i}{j}{|l],A}=2U^0\delta_{i[l}\delta_{k]j}+b_{kl}\m{i}{j}{1}-\m{i}{s}{[k|}\m{j}{s}{|l]}+\m{i}{j}{s}\m{s}{[k}{l]} . \label{11p_2} 
\eeqn
Note, in particular, that \eqref{X_U} and \eqref{xi_U} imply
\be
	U^0=\mbox{const} .
	\label{U0=const}
\ee

\section{Adapted coordinates and preferred frame}

\label{sec_adapted}

\subsection{Coordinates $(r,u)$ in the 2-plane spanned by $\bl$ and $\bn$ and scaling freedom}

\label{subsec_coord_u}

Recall that we have $\bl=\pa_r$ and $\bn=U\pa_r+X^A\pa_A$ (eq.~\eqref{frame_coefficients}). The commutator \eqref{comm_DelD} implied (first of \eqref{U}) that $X^A=X^{A0}$ is independent of $r$.  We thus have 
\be
 [\pa_r,X^A\pa_A]=0 ,
\ee
so that there exist a function $u$ that can be used as a coordinate together with $r$ and such that $X^A\pa_A=\pa_u$. In other words, we have now a coordinate system $(r,x^A)$=$(r,u,x^\alpha)$ (the ($n-2$) coordinates $x^\alpha$ will be specified in the following) such that
\be
 X^A=\delta^A_ u , 
\ee
i.e., 
\be
 \bn=U\pa_r+\pa_u .
 \label{n_adapted}
\ee

Eqs.~\eqref{X_om} (with \eqref{U0=const}), \eqref{X_xi}, \eqref{xi_xi}, \eqref{X_B} and \eqref{11o_2} thus simplify to 
\beqn
 & & \omega^0_{i,u}=-\m{j}{i}{1}\omega^0_j ,  \label{omega_u} \\
 & & \xi^{A0}_{i,u}=-\m{j}{i}{1}\xi^{A0}_j , \label{xi_u} \\
 & & \xi^{B0}_{[i}\xi^{u0}_{j],B}=b_{ij}+\xi^{u0}_{k}\m{k}{[i}{j]} ,  \qquad \xi^{B0}_{[i}\xi^{\a 0}_{j],B}=\xi^{\a 0}_{k}\m{k}{[i}{j]} ,  \label{xi_xi_2} \\
 & & b_{ij,u}=2b_{k[i}\m{k}{j]}{1} , \label{B_u} \\
 & & \m{i}{j}{k,u}-\xi^{A0}_{k}\m{i}{j}{1,A}=-2\m{s}{[i|}{1}\m{s}{|j]}{k}-\m{i}{j}{s}\m{s}{k}{1} . \label{mijk_u}
\eeqn

Let us further note that a coordinate transformation  
	\be
	  r'=\lambda^{-1}r , \qquad u'=\lambda u ,  
		\label{rescaling}
	\ee
where $\lambda\neq0$ is a constant, accompanied by a boost
\be
 \bl'=\lambda\bl=\pa_{r'} , \qquad \bn'=\lambda^{-1}\bn=\lambda^{-2}U\pa_{r'}+\pa_{u'} ,
 \label{rescaling_boost}
\ee
produces the following rescaling (cf.~\eqref{L_def}, \eqref{inverse}, \eqref{der_asym})
\be
 b_{ij}'=\lambda^{-1}b_{ij}, \qquad {U^0}'=\lambda^{-2}U^0 , \qquad \Phi_0'=\lambda^{1-n}\Phi_0 , \qquad {\omega^0_i}'=\lambda^{-2}\omega^0_i , \qquad {\xi^{u 0}_{i}}'=\xi^{u 0}_{i} , \qquad {\xi^{\a 0}_{i}}'=\lambda^{-1}\xi^{\a 0}_{i} .
 \label{rescaling_param}
\ee 
This freedom will be useful in the following.

\subsection{Choice of the vectors $\bm_i$}

The results presented so far hold in any number of dimensions $n>4$. However, {\em from now on, we restrict ourselves to the case $n=6$}.

In order to conveniently specify the spatial part of our frame, we shall use the remaining freedom of $r$-independent spatial rotations (``spins''), i.e., $\bm^{i}\mapsto X^{i}_{\ j} \bm^{j}$ with $DX^{i}_{\ j}=0$ (the $X^{i}_{\ j}$ are $4\times4$ orthogonal matrices). Under these one has\footnote{Eqs.~\eqref{b_spin} and \eqref{mij1_spin} follow from the transformation properties of the Ricci rotation coefficients given in \cite{OrtPraPra07}, which also ensure that $r$-independent spins are indeed compatible with the various frame choices made previously.}
\beqn
 & & b_{ij}\mapsto X^{i}_{\ k}X^{j}_{\ l}b_{kl} , \qquad \label{b_spin} \\
 & &  \m{i}{j}{1}\mapsto X^{i}_{\ k}X^{j}_{\ l}\m{k}{l}{1}+X^{j}_{\ k} X^{i}_{\ k,u} . \label{mij1_spin}
\eeqn
(Note that $\m{i}{j}{1}$ does {\em not} transform homogeneously under an $r$-independent but $u$-dependent spin.) Now, thanks to~\eqref{b_spin}, we can adapt the spacelike frame vectors to an ``eigenframe'' of the antisymmetric matrix $b_{ij}$ (cf., e.g., cap.~IX of \cite{Gantmacherbook}), i.e., without loss of generality from now on we can take
\beqn
 b = {\rm diag}\left(\left[\begin{array}{cc} 0 & b_{23} \\ 
   -b_{23} & 0 \\
  \end {array}
 \right] ,
\left[\begin {array}{cc} 0 & b_{45} \\ 
   -b_{45} & 0 \\ \end {array}
 \right]\right) .
 \label{b_eigenf}
\eeqn
Then the l.h.s. and the r.h.s. of \eqref{B_u} must vanish separately, i.e.,
\be
 b_{ij,u}=0 , \qquad b_{k[i}\m{k}{j]}{1}=0 \qquad \mbox{(when~\eqref{b_eigenf} holds)} . \label{BM}
\ee

We have now to distinguish between two different cases:
\begin{enumerate}[(i)]
 \item\label{b_generic} $b_{45}\neq\pm b_{23}$ (``generic case''): from the second of~\eqref{BM} it follows immediately that 
	\be
		\m{2}{4}{1}=\m{2}{5}{1}=\m{3}{4}{1}=\m{3}{5}{1}=0 ,
		\label{m1_generic}
	\ee
	i.e., $\m{2}{3}{1}$ and $\m{4}{5}{1}$ are the only non-zero components of $\m{i}{j}{1}$. However, a spin in the plane $(23)$ by an angle $\theta$, while preserving \eqref{b_eigenf}, produces the transformation (cf. \eqref{mij1_spin})
	\be
		\m{2}{3}{1}\mapsto \m{2}{3}{1}+\theta_{,u} ,
  \ee		
	which can be used to arrive at $\m{2}{3}{1}=0$ (the remaining $\m{i}{j}{1}$ are unchanged thanks to~\eqref{m1_generic}). Similarly, a spin in the plane $(45)$ can be used to set $\m{4}{5}{1}=0$, so that in the frame in use we finally have $\m{i}{j}{1}=0$.

 \item\label{b_deg} $b_{45}=b_{23}(\neq0)$: due to the degeneracy in the eigenvalues of $b$, one can now only conclude that
	\be
		\m{2}{4}{1}=\m{3}{5}{1} , \qquad \m{2}{5}{1}=-\m{3}{4}{1} .
	\ee
		On the other hand, here the canonical form~\eqref{b_eigenf} with $b_{45}=b_{23}$ is invariant under a larger set of spins,\footnote{Namely, a spin (25) followed by a spin (34) with an opposite angle, and a spin (24) followed by a spin (35) with the same angle.} which can be used to arrive again at~\eqref{m1_generic} -- one can then proceed as in case~\eqref{b_generic} to set $\m{i}{j}{1}=0$ in a suitable eigenframe of $b$. Let us also observe that the present case coincide with $\bl$ being {\em shearfree} and twisting (and thus necessarily expanding \cite{OrtPraPra07}), since here $b^2=f_0I$ (with $f_0=-b_{23}^2=-b_{45}^2$), which with \eqref{inverse} implies
\be
	L=\frac{1}{r^2-f_0}(rI+b) .
	\label{L_shearfree}
\ee		
(The case $b_{45}=-b_{23}$ corresponds to simply relabeling the frame vectors and need not be discussed separately). 
\end{enumerate}

Without loosing generality, from now on we can thus employ a parallelly transported frame such that, in both cases~\eqref{b_generic} and \eqref{b_deg},
\be
 \m{i}{j}{1}=0 .
 \label{mij1=0}
\ee

With~\eqref{omega_u} and \eqref{xi_u}, this gives
\be
	\omega^0_{i,u}=0 , \qquad \xi^{A0}_{i,u}=0 , \label{om_xi_u}
\ee	
while eqs.~\eqref{mijk_u} and \eqref{11p_2} reduce to
\beqn
 & & \m{i}{j}{k,u}=0 , \\
 & & \xi^{A0}_{[k|}\m{i}{j}{|l],A}=2U^0\delta_{i[l}\delta_{k]j}-\m{i}{s}{[k|}\m{j}{s}{|l]}+\m{i}{j}{s}\m{s}{[k}{l]} . \label{11p_3}
\eeqn

Let us now consider eq.~\eqref{xi_B}. Using~\eqref{b_eigenf}, this can be written explicitly as
\beqn
	& & \omega^0_2=\xi^{A0}_{3}b_{23,A}=b_{23}\m{3}{4}{4}+b_{45}\m{2}{5}{4}=b_{23}\m{3}{5}{5}-b_{45}\m{2}{4}{5}  , \label{om2} \\
	& & \omega^0_3=-\xi^{A0}_{2}b_{23,A}=-b_{23}\m{2}{4}{4}+b_{45}\m{3}{5}{4}=-b_{23}\m{2}{5}{5}-b_{45}\m{3}{4}{5}  , \\
	& & \omega^0_4=\xi^{A0}_{5}b_{45,A}=-b_{23}\m{3}{4}{2}-b_{45}\m{2}{5}{2}=b_{23}\m{2}{4}{3}-b_{45}\m{3}{5}{3} , \\
	& & \omega^0_5=-\xi^{A0}_{4}b_{45,A}=-b_{23}\m{3}{5}{2}+b_{45}\m{2}{4}{2}=b_{23}\m{2}{5}{3}+b_{45}\m{3}{4}{3} , \\ 
	& & \xi^{A0}_{4}b_{23,A}=0=\xi^{A0}_{5}b_{23,A} , \qquad \xi^{A0}_{2}b_{45,A}=0=\xi^{A0}_{3}b_{45,A} , \label{b_der} \\ 
	& & b_{45}\m{2}{5}{3}=-b_{23}\m{3}{4}{3} , \qquad b_{45}\m{3}{5}{2}=b_{23}\m{2}{4}{2} , \qquad b_{45}\m{2}{4}{3}=b_{23}\m{3}{5}{3} , \qquad b_{45}\m{3}{4}{2}=-b_{23}\m{2}{5}{2} , \label{B_m_243} \\
	& & b_{45}\m{2}{5}{5}=-b_{23}\m{3}{4}{5} , \qquad b_{45}\m{2}{4}{4}=b_{23}\m{3}{5}{4} , \qquad b_{45}\m{3}{5}{5}=b_{23}\m{2}{4}{5} , \qquad b_{45}\m{3}{4}{4}=-b_{23}\m{2}{5}{4} . \label{B_m_245} 
\eeqn

Further consequences of \eqref{om2}--\eqref{B_m_243}, however, need again to be studied separately in the two possible cases~\eqref{b_generic} (section~\ref{sec_generic}) and \eqref{b_deg} (section~\ref{sec_shearfree}) defined above.

\section{Complete integration: generic case ($b_{45}\neq\pm b_{23}$)}

\label{sec_generic}

We still have a residual freedom of $r$- and $u$-independent spins in the planes $(23)$ and $(45)$. This leaves \eqref{b_eigenf} and \eqref{mij1=0} unchanged, while 
\be
 \omega^0_i\mapsto X^{i}_{\ j}\omega^0_j .
\ee
Since $\omega^0_{i,r}=0=\omega^0_{i,u}$, this can be used to set
\be
 \omega^0_3=0 , \qquad \omega^0_5=0 . \label{om3_5=0}
\ee
With this choice, and assuming $b_{45}\neq0$ (without loosing generality since we have $b_{ij}\neq0$), eqs.~\eqref{om2}--\eqref{B_m_245} reduce to
\beqn
	& & \omega^0_2=\xi^{A0}_{3}b_{23,A}=\frac{b_{23}^2-b_{45}^2}{b_{45}}\m{2}{4}{5} , \qquad \xi^{A0}_{2}b_{23,A}=\xi^{A0}_{4}b_{23,A}=\xi^{A0}_{5}b_{23,A}=0 , \label{xi_b23} \\
	& & \omega^0_4=\xi^{A0}_{5}b_{45,A}=\frac{b_{23}^2-b_{45}^2}{b_{45}}\m{2}{5}{2} , \qquad \xi^{A0}_{4}b_{45,A}=\xi^{A0}_{2}b_{45,A}=\xi^{A0}_{3}b_{45,A}=0 , \label{xi_b45} \\ 
	& & \m{3}{5}{3}=\m{2}{5}{2} , \qquad \m{2}{5}{4}=-\m{2}{4}{5} , \qquad -\m{3}{4}{2}=\m{2}{4}{3}=\frac{b_{23}}{b_{45}}\m{2}{5}{2} , \qquad  \m{3}{5}{5}=\m{3}{4}{4}=\frac{b_{23}}{b_{45}}\m{2}{4}{5} , \label{m_generic} \\
	& & \m{3}{5}{2}=\m{2}{4}{2}=\m{2}{5}{3}=\m{3}{4}{3}=\m{2}{4}{4}=\m{3}{5}{4}=\m{2}{5}{5}=\m{3}{4}{5}=0 . \label{m_generic2} 
\eeqn

Since $b_{ij,r}=0=b_{ij,u}$, eqs.~\eqref{xi_b23}, \eqref{xi_b45} mean that 
	\be
	  b_{23}=\mbox{const} \Leftrightarrow \omega^0_2=0 , \qquad  b_{45}=\mbox{const} \Leftrightarrow \omega^0_4=0 .
		\label{b_const}
	\ee

Next, \eqref{xi_om} together with \eqref{xi_om_a} leads to (using \eqref{om3_5=0}, \eqref{m_generic} and \eqref{m_generic2})
\beqn
	& & \xi^{A0}_{3}\omega^0_{2,A}=-\omega^0_{2}\m{2}{3}{2} , \qquad \xi^{A0}_{5}\omega^0_{2,A}=\omega^0_{4}\m{2}{4}{5} , \qquad \xi^{A0}_{2}\omega^0_{2,A}=0=\xi^{A0}_{4}\omega^0_{2,A} ,  \label{xi_om1} \\
	& & \xi^{A0}_{5}\omega^0_{4,A}=-\omega^0_{4}\m{4}{5}{4} , \qquad \xi^{A0}_{3}\omega^0_{4,A}=-\omega^0_{2}\m{2}{4}{3} , \qquad \xi^{A0}_{4}\omega^0_{4,A}=0=\xi^{A0}_{2}\omega^0_{4,A} , \label{xi_om2} \\
	& & \omega^0_2\m{2}{3}{3}=0 , \qquad \omega^0_2\m{2}{3}{5}=0 , \qquad \omega^0_4\m{4}{5}{5}=0 , \qquad \omega^0_4\m{4}{5}{3}=0 , \label{xi_om3} \\ 
	& & \omega^0_2\m{2}{5}{2}+\omega^0_4\m{4}{5}{2}=0 , \qquad \omega^0_2\m{2}{3}{4}-\omega^0_4\m{3}{4}{4}=0 , \label{xi_om4} \\
	& & 2U^0b_{23}=\omega^0_2\m{2}{3}{2}+\omega^0_4\m{2}{4}{3} , \qquad 2U^0b_{45}=\omega^0_4\m{4}{5}{4}-\omega^0_2\m{2}{4}{5} . \label{xi_om5}
\eeqn

Let us now consider the second of~\eqref{xi_xi_2}, which, thanks to~\eqref{om_xi_u}, reduces to $\xi^{\b0}_{[i}\xi^{\a 0}_{j],\b}=\xi^{\a 0}_{k}\m{k}{[i}{j]}$. Using~\eqref{m_generic} and \eqref{m_generic2}, this reads
\beqn
 & & 2\xi^{\b0}_{[2}\xi^{\a 0}_{3],\b}=-\xi^{\a 0}_{2}\m{2}{3}{2}-\xi^{\a 0}_{3}\m{2}{3}{3}-2\xi^{\a 0}_{4}\m{2}{4}{3} , \label{xi_xi1} \\
 & & 2\xi^{\b0}_{[2}\xi^{\a 0}_{4],\b}=2\xi^{\a 0}_{3}\m{3}{[2}{4]}+2\xi^{\a 0}_{5}\m{5}{[2}{4]} , \\
 & & 2\xi^{\b0}_{[2}\xi^{\a 0}_{5],\b}=-\xi^{\a 0}_{2}\m{2}{5}{2}-\xi^{\a 0}_{3}\m{2}{3}{5}+2\xi^{\a 0}_{4}\m{4}{[2}{5]} , \\
 & & 2\xi^{\b0}_{[4}\xi^{\a 0}_{3],\b}=\xi^{\a 0}_{4}\m{3}{4}{4}-\xi^{\a 0}_{5}\m{4}{5}{3}+2\xi^{\a 0}_{2}\m{2}{[4}{3]} , \\
 & & 2\xi^{\b0}_{[3}\xi^{\a 0}_{5],\b}=\xi^{\a 0}_{2}\m{2}{3}{5}-\xi^{\a 0}_{3}\m{3}{5}{3}-\xi^{\a 0}_{4}\m{4}{5}{3}-\xi^{\a 0}_{5}\m{3}{5}{5} , \\
 & & 2\xi^{\b0}_{[4}\xi^{\a 0}_{5],\b}=2\xi^{\a 0}_{2}\m{2}{4}{5}-\xi^{\a 0}_{4}\m{4}{5}{4}-\xi^{\a 0}_{5}\m{4}{5}{5} . \label{xi_xi6}
\eeqn

Applying these to $b_{23,\a}$ and $b_{45,\a}$ (using \eqref{xi_b23}, \eqref{xi_b45}, \eqref{xi_om1}--\eqref{xi_om3}) one further obtains
\beqn
	& & \omega^0_2\m{3}{[2}{4]}=0 , \qquad \omega^0_2\m{2}{[3}{4]}=0 , \label{om2_m} \\ 
	& & \omega^0_4\m{5}{[2}{4]}=0 , \qquad \omega^0_4\m{4}{[2}{5]}=0 . \label{om4_m} 
\eeqn	
(These are not all independent due to \eqref{m_generic}.)

For further analysis it will be necessary to distinguish among various subcases, depending on the possible vanishing of  $\omega^0_2$ and $\omega^0_4$.

\subsection{Case $\d b_{23}\neq0\neq\d b_{45}$: general doubly-spinning Myers-Perry metric}

Since neither $b_{23}$ nor $b_{45}$ are constant (thus, in particular, $b_{23}\neq0\neq b_{45}$), because of \eqref{b_const} here we have $\omega^0_2\neq0\neq\omega^0_4$. Eqs.~\eqref{xi_om3}, \eqref{om2_m} and \eqref{om4_m} thus give
\beqn
  & & \m{2}{3}{3}=\m{2}{3}{5}=\m{4}{5}{5}=\m{4}{5}{3}=0 , \label{m_gen3} \\ 
	& & \m{3}{[2}{4]}=\m{2}{[3}{4]}=\m{5}{[2}{4]}=\m{4}{[2}{5]}=0 , \label{m_gen4}
\eeqn	
thanks to which \eqref{xi_xi1}--\eqref{xi_xi6} reduce to (recall also \eqref{m_generic})
\beqn
 & & 2\xi^{\b0}_{[2}\xi^{\a 0}_{3],\b}=-\xi^{\a 0}_{2}\m{2}{3}{2}-2\xi^{\a 0}_{4}\m{2}{3}{4} , \qquad 2\xi^{\b0}_{[2}\xi^{\a 0}_{4],\b}=0 , \qquad 2\xi^{\b0}_{[2}\xi^{\a 0}_{5],\b}=-\xi^{\a 0}_{2}\m{2}{5}{2} , \label{xi_xia} \\
 & & 2\xi^{\b0}_{[4}\xi^{\a 0}_{3],\b}=\xi^{\a 0}_{4}\m{3}{4}{4} , \qquad  2\xi^{\b0}_{[3}\xi^{\a 0}_{5],\b}=-\xi^{\a 0}_{3}\m{2}{5}{2}-\xi^{\a 0}_{5}\m{3}{4}{4} , \qquad 2\xi^{\b0}_{[4}\xi^{\a 0}_{5],\b}=2\xi^{\a 0}_{2}\m{2}{4}{5}-\xi^{\a 0}_{4}\m{4}{5}{4} . \label{xi_xib}
\eeqn

These relations suggest how to define commuting vector fields $\hx_2$, $\hx_3$, $\hx_4$ and $\hx_5$ spanning the subspace of the $x^\a$. Namely, choosing 
\beqn
 & & \hx_2\equiv\left[\omega^0_2(\alpha b_{45}^2+\beta)\xi^{\a 0}_{2}+\omega^0_4(\alpha b_{23}^2+\beta)\xi^{\a 0}_{4}\right]\pa_\a , \qquad \hx_3\equiv\frac{1}{\omega^0_2}\xi^{\a 0}_{3}\pa_\a , \label{hx1} \\
 & & \hx_4\equiv\left[\omega^0_2(\gamma b_{45}^2+\delta)\xi^{\a 0}_{2}+\omega^0_4(\gamma b_{23}^2+\delta)\xi^{\a 0}_{4}\right]\pa_\a , \qquad \hx_5\equiv\frac{1}{\omega^0_4}\xi^{\a 0}_{5}\pa_\a , \label{hx2} 
\eeqn
where $\alpha,\beta,\gamma,\delta$ are arbitrary constants such that $\alpha\delta-\beta\gamma\neq0$ (a convenient choice will be specified later), one can use \eqref{xi_xia}, \eqref{xi_xib} with \eqref{xi_b23}--\eqref{m_generic}, \eqref{xi_om1}, \eqref{xi_om2} and \eqref{xi_om4} to verify that $[\hx_i,\hx_j]=0$ and $[\hx_i,\pa_r]=0=[\hx_i,\pa_u]$ for any choice of $i,j=2,3,4,5$.  One can thus define the adapted ``transverse'' coordinates $x^\a=(y_1,\phi_1,y_2,\phi_2)$ via
\be
  \pa_{\phi_1}\equiv \hx_2 , \qquad \pa_{y_1}\equiv \hx_3 , \qquad \pa_{\phi_2}\equiv \hx_4 , \qquad \pa_{y_2}\equiv \hx_5 . 
	\label{phi_y}
\ee
With these, by \eqref{xi_b23} and \eqref{xi_b45} one has $b_{23,y_1}=1=b_{45,y_2}$, $b_{23,y_2}=b_{23,\phi_1}=b_{23,\phi_2}=0$, $b_{45,y_1}=b_{45,\phi_1}=b_{45,\phi_2}=0$, so that one can always choose $(y_1,y_2)$ such that
\be
 y_1\equiv b_{23} , \qquad y_2\equiv b_{45} .
 \label{y1_y2_eigenval}
\ee
(The coordinates $\phi_1$ and $\phi_2$ can be combined linearly in an arbitrary way, amounting to a redefinition of $\alpha,\beta,\gamma,\delta$ in \eqref{hx1}, \eqref{hx2}.)

In these coordinates, we can integrate \eqref{xi_om1} and \eqref{xi_om2} using \eqref{xi_om5} (and a subset of \eqref{xi_b23}--\eqref{m_generic}) to obtain the explicit form of $\omega^0_2$ and $\omega^0_4$ (up to a sign), namely,
\be
	(\omega^0_2)^2=\frac{2U^0y_1^4-c_0y_1^2-d_0}{y_2^2-y_1^2} , \qquad (\omega^0_4)^2=\frac{2U^0y_2^4-c_0y_2^2-d_0}{y_1^2-y_2^2} , 
	\label{om_2_4}
\ee
where $c_0$ and $d_0$ are integration constants (such that $(\omega^0_2)^2>0$, $(\omega^0_4)^2>0$ for a suitable range of $(y_1,y_2)$).\footnote{All the $\m{i}{j}{k}$ are thus now also determined. In particular, for later calculations it is useful to observe that (cf. \eqref{xi_om1}, \eqref{xi_om2}) $\m{2}{3}{2}=-\omega^0_{2,y_1}$, $\m{2}{4}{5}=\omega^0_{2,y_2}$, $\m{2}{4}{3}=-\omega^0_{4,y_1}$ and $\m{4}{5}{4}=-\omega^0_{4,y_2}$.} Furthermore, inverting \eqref{hx1}, \eqref{hx2} (with \eqref{phi_y}) gives the components $\xi^{\a 0}_{i}$, i.e.,
\beqn
 & & \xi^{\a 0}_{2}\pa_\a=\frac{\omega^0_2(\beta\gamma-\alpha\delta)^{-1}}{2U^0y_1^4-c_0y_1^2-d_0}\left[-(\gamma y_1^2+\delta)\pa_{\phi_1}+(\alpha y_1^2+\beta)\pa_{\phi_2}\right] , \label{xi2} \\
 & & \xi^{\a 0}_{4}\pa_\a=\frac{\omega^0_4(\beta\gamma-\alpha\delta)^{-1}}{2U^0y_2^4-c_0y_2^2-d_0}\left[-(\gamma y_2^2+\delta)\pa_{\phi_1}+(\alpha y_2^2+\beta)\pa_{\phi_2}\right] , \label{xi4} \\
 & & \xi^{\a 0}_{3}\pa_\a=\omega^0_2\pa_{y_1} , \qquad \xi^{\a 0}_{5}\pa_\a=\omega^0_4\pa_{y_2} . \label{xi3_5}
\eeqn

Next, we can determine the remaining components $\xi^{u 0}_{i}$ using the first of~\eqref{xi_xi_2} (with \eqref{b_eigenf}, \eqref{m_generic}, \eqref{m_generic2}, \eqref{m_gen3} and \eqref{m_gen4}). Let us first consider its $ij=35$ component, which gives
\be
 \omega^0_2\xi^{u 0}_{5,y_1}-\omega^0_4\xi^{u 0}_{3,y_2}=-\xi^{u 0}_{3}\m{2}{5}{2}-\xi^{u 0}_{5}\m{3}{4}{4} .
\ee 
Thanks to \eqref{xi_om1}, \eqref{xi_om2}, \eqref{m_gen4} and  \eqref{xi_om4}, this can be rewritten as
\be
 \left(\frac{\xi^{u 0}_{3}}{\omega^0_2}\right)_{,y_2}=\left(\frac{\xi^{u 0}_{5}}{\omega^0_4}\right)_{,y_1} .
\label{u0_integrability}
\ee 

This can be used to show that, without loss of generality, we can always set $\xi^{u 0}_{3}=0=\xi^{u 0}_{5}$. Namely, under a coordinate transformation 
\be
 u\mapsto u+V(y_1,y_2,\phi_1,\phi_2) ,
 \label{u_transf}
\ee
one has (with \eqref{xi3_5})
\be
  \xi^{u 0}_{2}\mapsto \xi^{u 0}_{2}+\xi^{\a 0}_2V_{,\a} , \qquad \xi^{u 0}_{3}\mapsto \xi^{u 0}_{3}+\omega^0_2V_{,y_1} , \qquad \xi^{u 0}_{4}\mapsto \xi^{u 0}_{4}+\xi^{\a 0}_4V_{,\a} , \qquad \xi^{u 0}_{5}\mapsto \xi^{u 0}_{5}+\omega^0_4V_{,y_2} .
	\label{u0}
\ee
(The $\omega^0_i$, $\xi^{\a 0}_{i}$ and $X^A=\delta^A_u$ are unchanged.) One can thus choose $V$ so that, simultaneously, $\xi^{u 0}_{3}+\omega^0_2V_{,y_1}=0=\xi^{u 0}_{5}+\omega^0_4V_{,y_2}$ in \eqref{u0}, since the corresponding integrability condition is clearly \eqref{u0_integrability} and thus identically satisfied. After the transformation \eqref{u_transf}, from now on we shall thus have
\be
 \xi^{u 0}_{3}=0 , \qquad \xi^{u 0}_{5}=0 .
\label{xiu3_5=0}
\ee

This choice simplifies the remaining components of the first of~\eqref{xi_xi_2}. The components $ij=25,23,34,45$ now take the form 
\beqn
	& & \xi^{u 0}_{2,y_2}=-\frac{y_2}{y_2^2-y_1^2}\xi^{u 0}_{2} , \qquad -(\omega^0_2\xi^{u 0}_{2})_{,y_1}=2y_1+2\xi^{u 0}_{4}\omega^0_{4,y_1} , \\
  & & \xi^{u 0}_{4,y_1}=-\frac{y_1}{y_1^2-y_2^2}\xi^{u 0}_{4} , \qquad -(\omega^0_4\xi^{u 0}_{4})_{,y_2}=2y_2+2\xi^{u 0}_{2}\omega^0_{2,y_2} .
\eeqn
These can be integrated to obtain (recall \eqref{om_2_4})
\be
	\xi^{u 0}_2=\frac{y_1^4-e_0y_1^2-f_0}{2U^0y_1^4-c_0y_1^2-d_0}\omega^0_2 , \qquad \xi^{u 0}_4=\frac{y_2^4-e_0y_2^2-f_0}{2U^0y_2^4-c_0y_2^2-d_0}\omega^0_4 , 
 \label{xiu_2_4}
\ee
where $e_0=e_0(\phi_1,\phi_2)$ and $f_0=f_0(\phi_1,\phi_2)$ are integration functions. Using \eqref{xiu_2_4} and \eqref{xi2}, \eqref{xi4}, the component $ij=24$ of the first of~\eqref{xi_xi_2} becomes a constraint on $e_0$ and $f_0$, i.e.,
\be
 (e_0\delta-f_0\gamma)_{,\phi_1}=(e_0\beta-f_0\alpha)_{,\phi_2} .
 \label{u0_integrability2}
\ee

Under a coordinate transformation \eqref{u_transf}, \eqref{u0} with $V_{,y_1}=0=V_{,y_2}$ (which preserves~\eqref{xiu3_5=0}), one finds (using \eqref{xi2}, \eqref{xi4}) that in \eqref{xiu_2_4}
\be
  e_0\mapsto e_0+\frac{\gamma}{\beta\gamma-\alpha\delta}V_{,\phi_1}-\frac{\alpha}{\beta\gamma-\alpha\delta}V_{,\phi_2} , \qquad f_0\mapsto f_0+\frac{\delta}{\beta\gamma-\alpha\delta}V_{,\phi_1}-\frac{\beta}{\beta\gamma-\alpha\delta}V_{,\phi_2} . 
	\label{rescal_e0_f0}
\ee
This can be used to assign to $e_0$ and $f_0$ an arbitrary {\em constant} value, since \eqref{u0_integrability2} makes sure that the corresponding integrability condition is satisfied. A convenient choice of these constants will differ in the two possible cases $U^0\neq0$ and $U^0=0$, as we now discuss.

\subsubsection{Subcase $U^0\neq0$}

First, let us use the freedom \eqref{rescal_e0_f0} to choose $e_0$, $f_0$ in \eqref{xiu_2_4} such that $2U^0(s^4-e_0s^2-f_0)=2U^0s^4-c_0s^2-d_0$, so that 
\be
 2U^0\xi^{u 0}_2=\omega^0_2 , \qquad 2U^0\xi^{u 0}_4=\omega^0_4 .
\ee
Next, since both functions in \eqref{om_2_4} must be strictly positive, it follows that $c_0$ and $d_0$ must be such that the polynomial appearing in the numerators of \eqref{om_2_4} can be factorized as 
\be 
 2U^0s^4-c_0s^2-d_0=2\e|U^0|(s^2-s_1)(s^2-s_2) ,
 \label{s1_s2}
\ee
(with $s_1\neq s_2$ and at least one of $s_1$, $s_2$ strictly positive for $\e=-1$, both strictly positive for $\e=+1$), which we will use henceforth, and where we have defined
\be
	\e\equiv\mbox{sign}(U^0)=\pm 1 . 
	\label{def_eps}
\ee

Further, using the scaling freedom~\eqref{rescaling_param} accompanied by an additional parameter redefinition
\beqn
 & & s_1'=\frac{s_1}{\lambda^2}, \quad s_2'=\frac{s_2}{\lambda^2} , \qquad \a'=\lambda^5\a , \qquad \ng'=\lambda^5\ng , \qquad \nb'=\lambda^3\nb , \qquad \nd'=\lambda^3\nd ,
\eeqn
with
\be
 \lambda=\sqrt{2|U^0|} ,
\ee
we can set $2{U^0}'=\e$. Apart from this simplification, the expressions \eqref{om_2_4}, \eqref{xi2}--\eqref{xi3_5} retain their form, with the factorization \eqref{s1_s2}. Finally, if we choose $\a'=(s_1'-s_2')^{-1}$, $\ng'=(s_2'-s_1')^{-1}$, $\nb'=s_1(s_2'-s_1')^{-1}$, $\nd'=s_2(s_1'-s_2')^{-1}$, we arrive at the following simplified expressions for the asymptotic quantities  (all primes will be dropped hereafter) 
\beqn
 & & (\omega^0_2)^2=\e\frac{(s_1-y_1^2)(s_2-y_1^2)}{y_2^2-y_1^2} , \qquad (\omega^0_4)^2=\e\frac{(s_1-y_2^2)(s_2-y_2^2)}{y_1^2-y_2^2} , \qquad \omega^0_3=0=\omega^0_5 , \\
 & & \xi^{u 0}_2=\e\omega^0_2 , \qquad \xi^{u 0}_4=\e\omega^0_4 , \qquad \xi^{u 0}_3=0=\xi^{u 0}_5 , \\
 & & \xi^{\a 0}_{2}\pa_\a=-\e\omega^0_2\left(\frac{1}{s_1-y_1^2}\pa_{\phi_1}+\frac{1}{s_2-y_1^2}\pa_{\phi_2}\right) , \qquad \xi^{\a 0}_{3}\pa_\a=\omega^0_2\pa_{y_1} , \\
 & & \xi^{\a 0}_{4}\pa_\a=-\e\omega^0_4\left(\frac{1}{s_1-y_2^2}\pa_{\phi_1}+\frac{1}{s_2-y_2^2}\pa_{\phi_2}\right) , \qquad \xi^{\a 0}_{5}\pa_\a=\omega^0_4\pa_{y_2} , \\
 & & U^0=\frac{\epsilon}{2} , \qquad X^{A0}=\delta^A_u , 
\eeqn
with the conditions mentioned above on $s_1$ and $s_2$. This shows (recall the comments following \eqref{Cijkm_leading}) that the present solution describes the (generalized) {\em doubly-spinning Myers-Perry metric with unequal spins} of appendix~\ref{app_MP_generic} (using the fact that $\phi_1$ and $\phi_2$ can be multiplied by an arbitrary non-zero constant -- see also the comments following \eqref{g1_g2}).

\subsubsection{Subcase $U^0=0$}

\label{subsubsec_MP_U0=0}

In this case, we necessarily have $c_0>0$ and $d_0<0$ in \eqref{om_2_4}. Using~\eqref{rescal_e0_f0}, let us now make the simplifying choice $e_0=-d_0/c_0$, $f_0=0$ in \eqref{xiu_2_4}. Using the scaling freedom~\eqref{rescaling_param} similarly as above, and choosing conveniently the arbitrary constants $\a$, $\nb$, $\ng$, $\nd$ (we now skip these details and again drop the primes of the rescaled quantities), without loss of generality one can write the asymptotic quantities in the form
\beqn
 & & (\omega^0_2)^2=\frac{a^2-y_1^2}{y_2^2-y_1^2} , \qquad (\omega^0_4)^2=\frac{a^2-y_2^2}{y_1^2-y_2^2} , \qquad \omega^0_3=0=\omega^0_5 , \\
 & & \xi^{u 0}_2=- y_1^2\omega^0_2 , \qquad \xi^{u 0}_4=- y_2^2\omega^0_4 , \qquad \xi^{u 0}_3=0=\xi^{u 0}_5 , \\
 & & \xi^{\a 0}_{2}\pa_\a=-\omega^0_2\left(\pa_{\phi_1}+\frac{1}{y_1^2-a^2}\pa_{\phi_2}\right) , \qquad \xi^{\a 0}_{3}\pa_\a=\omega^0_2\pa_{y_1} , \\
 & & \xi^{\a 0}_{4}\pa_\a=-\omega^0_4\left(\pa_{\phi_1}+\frac{1}{y_2^2-a^2}\pa_{\phi_2}\right) , \qquad \xi^{\a 0}_{5}\pa_\a=\omega^0_4\pa_{y_2} , \\
 & & U^0=0 , \qquad X^{A0}=\delta^A_u , 
\eeqn
where the constant $a^2$ corresponds to a redefinition of the rescaled $d_0$.

It is then not difficult to determine the exact form of the coefficients of the frame vectors in \eqref{frame_coefficients} using \eqref{omegai} and \eqref{U} with \eqref{inverse} and \eqref{b_eigenf} (recall \eqref{y1_y2_eigenval}).\footnote{To be precise, the exact form of $U$ in \eqref{U}  is not yet known at this stage, since we have not explicitly determined $\Phi_{ij}$. This can be done \cite{OrtPraPra_prep}, but an alternative easy way to find $U$ consists in writing it in terms of an unknown function $\rho(r,y_1,y_2)$ \cite{OrtPraPra_prep} as $U=-\mu r\rho^{-2}/2$ (with $\mu=$const) and then solving the vacuum Einstein equations for the metric \eqref{generic_U0=0}, which gives \eqref{rho_generic_U0=0}. (The fact that $\Phi_{ij}$, and thus $U$, are independent of $(u,\phi_1,\phi_2)$ follows from \eqref{Phi_leading} and the comments following it.)} These determine the contravariant form of the metric, which can be inverted to obtain
\beqn
 \d s^2=2\d r\left[\d u+(a^2-y_1^2-y_2^2)\d\phi_1+(a^2-y_1^2)(a^2-y_2^2)\d\phi_2\right]+(r^2+y_1^2)\frac{y_2^2-y_1^2}{a^2-y_1^2}\d y_1^2+(r^2+y_2^2)\frac{y_1^2-y_2^2}{a^2-y_2^2}\d y_2^2 \nonumber \\
 {}+2\d u\d\phi_1+(r^2+a^2-y_1^2-y_2^2)\d\phi_1^2-(r^2+a^2)(a^2-y_1^2)(a^2-y_2^2)\d\phi_2^2 \nonumber \\
 {}+\frac{\mu r}{\rho^2}\left[\d u+(a^2-y_1^2-y_2^2)\d\phi_1+(a^2-y_1^2)(a^2-y_2^2)\d\phi_2\right]^2 ,
 \label{generic_U0=0}
\eeqn
with
\be
 \rho^2=(r^2+y_1^2)(r^2+y_2^2) . 
 \label{rho_generic_U0=0}
\ee 
This spacetime is manifestly of the Kerr-Schild form (being flat for $\mu=0$). It can be verified that the 4-spaces of constant $r$ and $u$ become flat asymptotically for $r\to\infty$.

Defining new coordinates $(t,\psi_1,\psi_2)$
\be
 \d u=\d t+\frac{r^2(r^2+a^2)}{r^2-\mu r+a^2}\d r , \qquad \d\phi_1=\d\psi_1-\frac{r^2+a^2}{r^2-\mu r+a^2}\d r , \qquad \d\phi_2=\d\psi_2+\frac{1}{r^2-\mu r+a^2}\d r ,
\ee 
one obtains an alternative Boyer-Lindquist form of the metric~\eqref{generic_U0=0}
\beqn
 \d s^2=\frac{\mu r}{\rho^2}\left[\d t+(a^2-y_1^2-y_2^2)\d\psi_1+(a^2-y_1^2)(a^2-y_2^2)\d\psi_2\right]^2+(r^2+y_1^2)\frac{y_2^2-y_1^2}{a^2-y_1^2}\d y_1^2+(r^2+y_2^2)\frac{y_1^2-y_2^2}{a^2-y_2^2}\d y_2^2 \nonumber \\
 {}+\frac{\rho^2}{r^2-\mu r+a^2}\d r^2+2\d t\d\psi_1+(r^2+a^2-y_1^2-y_2^2)\d\psi_1^2-(r^2+a^2)(a^2-y_1^2)(a^2-y_2^2)\d\psi_2^2  .
\eeqn

This metric is not included in the metrics of \cite{MyePer86} (where $U^0=1/2$), but is contained in (48) of \cite{CheLuPop06} (without using their (49), and choosing their parameters to be $g=L_1=L_2=C_0=0$, $C_1=1$, $C_2=a^2$ -- and up to appropriate linear redefinitions of the time and angles coordinates). The Kretschmann scalar diverges for $r^2+y_1^2=0$ or $r^2+y_2^2=0$.
In these coordinates, $\ell_{a}^\pm\d x^a=\d t\pm\rho^2(r^2-\mu r+a^2)^{-1}\d r+(a^2-y_1^2-y_2^2)\d\psi_1+(a^2-y_1^2)(a^2-y_2^2)\d\psi_2$ define the two multiple WANDs \cite{PraPraOrt07,OrtPraPra09} (so that the Weyl tensor is of type D).

\subsection{Case $\d b_{23}=0\neq\d b_{45}$: special Myers-Perry metrics} 

Here $b_{23}$ is a constant, but not $b_{45}$, so that $\omega^0_2=0\neq\omega^0_4$. Therefore \eqref{xi_b23} and \eqref{m_generic} give 
\be
 \m{2}{4}{5}=\m{2}{5}{4}=\m{3}{4}{4}=\m{3}{5}{5}=0 .
\ee
Eqs.~\eqref{xi_om3}, \eqref{xi_om4} and \eqref{xi_om5} (using also the third of \eqref{m_generic}) further give
\beqn
  & & \m{4}{5}{5}=\m{4}{5}{3}=\m{4}{5}{2}=0 , \\
	& & b_{23}(\omega^0_4)^2=2b_{23}U^0(b_{23}^2-b_{45}^2) , \label{constr_om_4}
\eeqn	
while  \eqref{om2_m} and \eqref{om4_m} are satisfied identically.

The components $ijkl=2345$ and $ijkl=2435$ of \eqref{11p_3} read
\be
 \xi^{\a 0}_{4}\m{2}{3}{5,\a}-\xi^{\a 0}_{5}\m{2}{3}{4,\a}=-\m{2}{3}{4}\m{4}{5}{4} , \qquad \xi^{\a 0}_{5}\m{2}{4}{3,\a}=\m{2}{4}{3}\m{2}{5}{2} ,
 \label{11p_integrability}
\ee 
where $\m{4}{5}{4}$, $\m{2}{4}{3}$ and $\m{2}{5}{2}$ can be obtained from \eqref{xi_om5}, \eqref{m_generic} and \eqref{xi_b45}.

Further, we observe that here we still have a freedom of $r$- and $u$-independent  (23)-spins (thanks to $\omega^0_2=0=\omega^0_3$). Under these, one has, in particular \cite{OrtPraPra07}
	\be
		\m{2}{3}{4}\mapsto \m{2}{3}{4}+\xi^{\a 0}_{4}\theta_{,\a} , \qquad \m{2}{3}{5}\mapsto \m{2}{3}{5}+\xi^{\a 0}_{5}\theta_{,\a} .
		\label{spin_23_m23k}
  \ee		
This can be used to set simultaneously
\be
 \m{2}{[3}{4]}=0, \qquad \m{2}{3}{5}=0 , 
\label{m235=0}
\ee	
since the integrability condition following from the two equations~\eqref{spin_23_m23k} is identically satisfied thanks to~\eqref{11p_integrability} with \eqref{xi_xi6}.

Thanks to the above simplifications, eqs.~\eqref{xi_xi1}--\eqref{xi_xi6} now reduce to (recall \eqref{m_generic})
\beqn
 & & 2\xi^{\b0}_{[2}\xi^{\a 0}_{3],\b}=-\xi^{\a 0}_{2}\m{2}{3}{2}-\xi^{\a 0}_{3}\m{2}{3}{3}-2\xi^{\a 0}_{4}\m{2}{4}{3} , \qquad 2\xi^{\b0}_{[2}\xi^{\a 0}_{4],\b}=0 , \qquad 2\xi^{\b0}_{[2}\xi^{\a 0}_{5],\b}=-\xi^{\a 0}_{2}\m{2}{5}{2} , \label{xi_xia_case2} \\
 & & 2\xi^{\b0}_{[3}\xi^{\a 0}_{4],\b}=0 , \qquad 2\xi^{\b0}_{[3}\xi^{\a 0}_{5],\b}=-\xi^{\a 0}_{3}\m{2}{5}{2} , \qquad 2\xi^{\b0}_{[4}\xi^{\a 0}_{5],\b}=-\xi^{\a 0}_{4}\m{4}{5}{4} . \label{xi_xib_case2}
\eeqn

This shows that the distribution $\{\xi^{\a 0}_{4}\pa_\a,\xi^{\a 0}_{5}\pa_\a\}$ is integrable. It is spanned by the two commuting vector fields (recall \eqref{xi_om2} with $\omega^0_2=0$)
\be
 \hx_4\equiv\omega^0_4\xi^{\a 0}_{4}\pa_\a , \qquad \hx_5\equiv\frac{1}{\omega^0_4}\xi^{\a 0}_{5}\pa_\a , 
 \label{hx45_case2}
\ee
which can thus be used to define two coordinates $(y_2,\phi_2)$ via
\be
  \pa_{\phi_2}\equiv \hx_4 , \qquad \pa_{y_2}\equiv \hx_5 . 
	 \label{phi2_y2_case2}
\ee
Here one can choose (cf. the first of \eqref{xi_b45})
\be
 y_2\equiv b_{45} .
 \label{y2_case_2}
\ee
Then \eqref{xi_om2} and \eqref{xi_om5} give
\be
 (\omega^0_4)^2=p_0-2U^0y_2^2 ,
 \label{om_4_case2}
\ee
where $p_0$ is a constant such that (cf. \eqref{constr_om_4})
\be
 b_{23}(p_0-2U^0b_{23}^2)=0 , \qquad p_0-2U^0y_2^2>0 .
 \label{constr_p0}
\ee
(This implies that when $b_{23}\neq0$ one has $(\omega^0_4)^2=2U^0(b_{23}^2-y_2^2)>0$ and therefore $U^0\neq0$.)

The remaining non-trivial components of \eqref{11p_3} now reduce to
\beqn
 & & \m{2}{3}{2,y_2}=\frac{y_2}{b_{23}^2-y_2^2}\m{2}{3}{2} , \qquad \m{2}{3}{3,y_2}=\frac{y_2}{b_{23}^2-y_2^2}\m{2}{3}{3} , \label{m232_y2_case2} \\
 & & \m{2}{3}{2,\phi_2}=0 , \qquad \m{2}{3}{3,\phi_2}=0 ,\\
 & & \xi^{\a0}_{3}\m{2}{3}{2,\a}-\xi^{\a0}_{2}\m{2}{3}{3,\a}=(\m{2}{3}{2})^2+(\m{2}{3}{3})^2+\frac{10U^0b_{23}^2-p_0}{b_{23}^2-y_2^2} . \label{const_curv}
\eeqn

Now, \eqref{xi_xia_case2} and \eqref{xi_xib_case2} show that also the distribution $\{\xi^{\a 0}_{2}\pa_\a,\xi^{\a 0}_{3}\pa_\a,\xi^{\a 0}_{4}\pa_\a\}$ is integrable. Furthermore, the vectors 
\be
 \hx_2\equiv\sqrt{|b_{23}^2-y_2^2|}\xi^{\a 0}_{2}\pa_\a , \qquad \hx_3\equiv\sqrt{|b_{23}^2-y_2^2|}\xi^{\a 0}_{3}\pa_\a ,
 \label{hx23_case2}
\ee
commute with $\hx_4$ and $\hx_5$, but not among themselves. We can thus complete our coordinate system $x^\a=(z,\bar z,y_2,\phi_2)$ by defining a pair of complex conjugate coordinates $(z,\bar z)$ such that (cf., e.g., (4.32) of \cite{Talbot69})
\be
 \hx_2+i\hx_3=P(\pa_z+Q\pa_{\phi_2}) ,
 \label{xi_23}
\ee
where $P=P(z,\bar z)$ and $Q=Q(z,\bar z)$ are complex functions.

Then, from \eqref{m232_y2_case2} one obtains
\be
 \m{2}{3}{2}+i\m{2}{3}{3}=\frac{D_0}{\sqrt{|b_{23}^2-y_2^2|}} ,
\label{m232_complex}
\ee 
where $D_0(z,\bar z)$ is a (complex) function of integration. Substituting \eqref{xi_23} and \eqref{m232_complex} into the first of \eqref{xi_xia_case2} gives
\beqn
 & & D_0=-i\frac{P\bar P_{,z}}{\bar P} , \label{D0} \\
 & & \bar Q_{,z}-Q_{,\bar z}=i\frac{4\e b_{23}}{P\bar P} , \label{eq_Q}
\eeqn
where the definition~\eqref{def_eps} is used,\footnote{Here for $b_{23}\neq0$ one has also $\e=\mbox{sign}(b_{23}^2-y_2^2)$, cf. \eqref{constr_om_4}, \eqref{y2_case_2}.}
while \eqref{const_curv} becomes (using also \eqref{D0})	
\beqn
	 P\bar P(\ln P\bar P)_{,z\bar z}=K ,
	\label{const_curv_2} 
\eeqn
with $K$ given by (recall \eqref{constr_p0})
\beqn
	& & K=2p_0 \qquad  \mbox{(if $b_{23}=0$)} , \\
	& & K=16|U^0|b_{23}^2>0 \qquad  \mbox{(if $b_{23}\neq 0$)} .
\eeqn

It is not difficult to see that \eqref{const_curv_2} is equivalent to requiring that the following auxiliary 2-dimensional metric 
\be
 \d s^2_{(2)}=\frac{2\d z\d\bar z}{P\bar P} ,
\label{2_metric}
\ee
has constant Gaussian curvature $K$. A redefinition of the coordinate $z\mapsto z'(z)$ thus always exists such that (cf., e.g., eq.~(2.55) of \cite{NewTamUnt63} or (7.2) of \cite{Talbot69})
\be
 P=\bar P=1+\frac{K}{2}z\bar z ,
 \label{P_case2}
\ee 
where we dropped the prime over $z$. (A (23)-spin  may also be necessary to preserve the form \eqref{xi_23} with $P$ real.)

One can now also solve \eqref{eq_Q} to find (up to using a coordinate transformation $\phi_2\mapsto\phi_2+Z(z,\bar z)$) \cite{deFGodRea15,NewTamUnt63}
\beqn
	& & Q=0 \qquad  \mbox{(if $b_{23}=0$)} , \label{Q_case2_i} \\
	& & Q=\frac{1}{4U^0b_{23}}\frac{i}{zP} \qquad  \mbox{(if $b_{23}\neq 0$)} . \label{Q_case2_ii}
\eeqn

The last step consists now in determining the components $\xi^{u 0}_{i}$. First, it is useful to observe that, under a coordinate transformation ($V$ is a real function)
\be
	u\mapsto u+V(z,\bar z,y_2,\phi_2) ,
	\label{u_transf_case2}
\ee
one has (using \eqref{hx45_case2}, \eqref{phi2_y2_case2}, \eqref{hx23_case2}, \eqref{xi_23})
\beqn
  & & \xi^{u 0}_{2}+i\xi^{u 0}_{3}\mapsto \xi^{u 0}_{2}+i\xi^{u 0}_{3}+\frac{P}{\sqrt{|b_{23}^2-y_2^2|}}(V_{,z}+QV_{,\phi_2}) ,  \label{V1_case2} \\
	& & \xi^{u 0}_{4}\mapsto \xi^{u 0}_{4}+\frac{1}{\omega^0_4}V_{,\phi_2} , \qquad \xi^{u 0}_{5}\mapsto \xi^{u 0}_{5}+\omega^0_4V_{,y_2} . \label{V2_case2}
\eeqn
Since $\omega^0_4\neq0$, one can always choose a function $V$ such as to set 
\be
 \xi^{u 0}_{5}=0 .
 \label{xiu_5=0_case2}
\ee

Using this simplification, the components $ij=25,35$ and $ij=45$ (recall \eqref{om_4_case2}) of the first of~\eqref{xi_xi_2} can be written as 
\beqn
 & & \left(\xi^{u 0}_{2}+i\xi^{u 0}_{3}\right)_{,y_2}=\frac{y_2}{b_{23}^2-y_2^2}\left(\xi^{u 0}_{2}+i\xi^{u 0}_{3}\right) , \label{25_case2} \\
 & & (\omega^0_4\xi^{u 0}_4)_{,y_2}=-2y_2 . \label{45_case2}
\eeqn
Eq.~\eqref{45_case2} immediately gives $\xi^{u 0}_4=\left[g_0(z,\bar z,\phi_2)-y_2^2\right]/\omega^0_4$, where $g_0$ is an integration function. 
However, one can use a transformation \eqref{u_transf_case2} with $V_{,y_2}=0$ (thus preserving \eqref{xiu_5=0_case2}) to assign an arbitrary {\em constant} value to $g_0$ (cf. \eqref{V2_case2}) -- for later convenience we choose to set $g_0=\frac{p_0}{2U^0}$ if $U^0\neq0$, and $g_0=0$ if $U^0=0$ (in which case $p_0>0$ by \eqref{om_4_case2}), so that 
\beqn
 & & \xi^{u 0}_4=\frac{1}{2U^0}\sqrt{p_0-2U^0y_2^2}  \qquad \mbox{if } U^0\neq0 \label{xiu_4=0_case2_U0} , \\
 & & \xi^{u 0}_4=-\frac{y_2^2}{\sqrt{p_0}}  \qquad \mbox{if } U^0=0 . \label{xiu_4=0_case2_U0=0}
\eeqn

The components $ij=24,34$ of the first of~\eqref{xi_xi_2} then simply give
\be
 \left(\xi^{u 0}_{2}+i\xi^{u 0}_{3}\right)_{,\phi_2}=0 ,
\ee
which with \eqref{25_case2} leads to
\be
 \xi^{u 0}_{2}+i\xi^{u 0}_{3}=\frac{P}{\sqrt{|b_{23}^2-y_2^2|}}h_0(z,\bar z) ,
 \label{xiu_23=0_case2}
\ee
where $h_0$ is an integration function. 

The component $ij=23$ reads (using \eqref{xiu_4=0_case2_U0}, \eqref{xiu_4=0_case2_U0=0}, \eqref{xiu_23=0_case2}, \eqref{m232_complex}, \eqref{D0})
\be
 \bar h_{0,z}-h_{0,\bar z}=0 .
\ee
This condition ensures that we can use a transformation \eqref{u_transf_case2} with $V_{,y_2}=0=V_{,\phi_2}$ (thus preserving \eqref{xiu_5=0_case2} and \eqref{xiu_4=0_case2_U0}, \eqref{xiu_4=0_case2_U0=0}) to set (cf. \eqref{V1_case2})
\be
 h_0=0 .
 \label{h0}
\ee

\subsubsection{Subcase $b_{23}=0$: Myers-Perry metric with a single spin, and a limit thereof}

\label{subsubsec_b23=0}

		\begin{enumerate}
			\item $U^0\neq0$: let us define a rescaled parameter $\tilde p_0=p_0/[4\e(U^0)^2]$, and perform a transformation \eqref{rescaling}--\eqref{rescaling_param} accompanied by a further coordinate rescaling 
		$(z,\bar z,y_2,\phi_2)\mapsto (\zeta,\bar\zeta,y_2,\chi)$ 
				\be
					\zeta=\lambda^2 z , \qquad \chi=\lambda^3 \phi_2 , 
				\ee	
				with
		\be
			\lambda=\sqrt{2|U^0|} .
			\ee


Dropping the primes in \eqref{rescaling}--\eqref{rescaling_param} and redefining
\be
 y_2\mapsto y,
\ee
the result for this case (cf. \eqref{hx45_case2}, \eqref{phi2_y2_case2}, \eqref{om_4_case2}, \eqref{hx23_case2}, \eqref{xi_23}, \eqref{P_case2}, \eqref{Q_case2_i}, \eqref{xiu_4=0_case2_U0}, \eqref{xiu_23=0_case2}, \eqref{h0}) can be thus summarized as follows (up to rescaling $\bm_2+i\bm_3\mapsto\mbox{sign}(y)(\bm_2+i\bm_3)$)
\beqn
 & & \omega^0_2=\omega^0_3=\omega^0_5=0 , \qquad \omega^0_4=\sqrt{\epsilon(\tilde p_0-y^2)} , \\
 & & \xi^{u 0}_{2}=\xi^{u 0}_{3}=\xi^{u 0}_{5}=0 , \qquad \xi^{u 0}_{4}=\epsilon \sqrt{\epsilon(\tilde p_0-y^2)} , \\
 & & \left(\xi^{\a 0}_{2}+i\xi^{\a 0}_{3}\right)\pa_\a=\frac{P}{y}\pa_\zeta , \\
 & & \xi^{\a 0}_{4}\pa_\a=\frac{1}{\sqrt{\epsilon(\tilde p_0-y^2)}}\pa_\chi , \qquad \xi^{\a 0}_{5}\pa_\a=\sqrt{\epsilon(\tilde p_0-y^2)}\pa_{y} , \\
 & & U^0=\frac{\epsilon}{2} , \qquad X^{A0}=\delta^A_u , 
\eeqn
where
\be
 \epsilon=\pm 1 , \qquad P=1+\e\tilde p_0\zeta\bar\zeta .
\ee 

It follows that this solution corresponds to the {\em Myers-Perry metric with a single spin} of appendix~\ref{subsec_MP_single} (after identifying $\tilde p_0=a^2$).

		\item $U^0=0$:\label{b23=0_case2} here necessarily $p_0>0$ (cf. \eqref{om_4_case2}). For simplicity, let us relabel
			\be
				y_2\mapsto y , \qquad \phi_2\mapsto \chi .
			\ee
The result for this case (cf. \eqref{hx45_case2}, \eqref{phi2_y2_case2}, \eqref{om_4_case2}, \eqref{hx23_case2}, \eqref{xi_23}, \eqref{P_case2}, \eqref{Q_case2_i}, \eqref{xiu_4=0_case2_U0=0}, \eqref{xiu_23=0_case2}, \eqref{h0}) can be thus summarized as follows (up to rescaling $\bm_2+i\bm_3\mapsto\mbox{sign}(y)(\bm_2+i\bm_3)$)
\beqn
 & & \omega^0_2=\omega^0_3=\omega^0_5=0 , \qquad \omega^0_4=\sqrt{p_0} , \label{EE1}\\
 & & \xi^{u 0}_{2}=\xi^{u 0}_{3}=\xi^{u 0}_{5}=0 , \qquad \xi^{u 0}_{4}=-y^2/\sqrt{p_0} , \\
 & & \left(\xi^{\a 0}_{2}+i\xi^{\a 0}_{3}\right)\pa_\a=\frac{P}{y}\pa_z , \\
 & & \xi^{\a 0}_{4}\pa_\a=(1/\sqrt{p_0})\pa_\chi , \qquad \xi^{\a 0}_{5}\pa_\a=\sqrt{p_0}\pa_{y} , \\
 & & U^0=0 , \qquad X^{A0}=\delta^A_u , \label{EE5}
\eeqn
where
\be
 P=1+p_0 z\bar z	\qquad (p_0>0) .
 \label{P_1spin_limit}
\ee

(The parameter $p_0$ could be normalized to 1 using \eqref{rescaling}--\eqref{rescaling_param} and a further rescaling of $(z,\bar z,y,\chi)$, but we prefer not to do so for dimensional reasons.) Similarly as in section~\ref{subsubsec_MP_U0=0}, one can use the above asymptotic quantities to arrive at the full metric as follows
\beqn
 \d s^2=2\d r\left(\d u+y^2\d\chi\right)-2p_0\d u\d\chi+(r^2+y^2)\frac{\d y^2}{p_0}+p_0(r^2-y^2)\d\chi^2+r^2y^2\frac{4\d z\d\bar z}{P^2}+\frac{\mu r}{\rho^2}\left(\d u+y^2\d\chi\right)^2 ,
	\label{strange}
\eeqn
with \eqref{P_1spin_limit} and 
\be
 \rho^2=r^2(r^2+y^2)  .
\ee
The 4-spaces of constant $r$ and $u$ become flat asymptotically for $r\to\infty$.

One can define new coordinates $(t,\psi)$ using
\be
 \d u=\d t+\frac{r^2}{p_0-\frac{\mu}{r}}\d r , \qquad \d \chi=\d\psi+\frac{1}{p_0-\frac{\mu}{r}}\d r ,
\ee
so that the metric~\eqref{strange} takes the Boyer-Lindquist form
\beqn
 \d s^2=\frac{1}{r^2+y^2}\left[-\left(p_0-\frac{\mu}{r}\right)\left(\d t+y^2\d\psi\right)^2+p_0\left(\d t-r^2\d\psi\right)^2\right]+(r^2+y^2)\left(\frac{\d r^2}{p_0-\frac{\mu}{r}}+\frac{\d y^2}{p_0}\right)+r^2y^2\frac{4\d z\d\bar z}{P^2} .
	\label{strange_stationary}
\eeqn
This is metric (2.33) of \cite{Klemm98} (specialized to six dimensions and with a vanishing cosmological constant). At $r=0$ there is a curvature singularity (where the Kretschmann scalar diverges), while $p_0r=\mu$ represents a Killing horizon of the Killing vector field $\mu^2\pa_t+p_0^2\pa_\psi$. In these coordinates, $\ell^{\pm}_a\d x^a=\d t\pm\frac{r(r^2+y^2)}{p_0r-\mu}\d r+y^2\d\psi$ define the two multiple WANDs \cite{PraPraOrt07,OrtPraPra09} (so that the Weyl tensor is of type D).

It may be interesting to observe that metric~\eqref{strange_stationary} can be obtained from the single-spin Myers-Perry metric \eqref{MP_6D_single_stat_2} (dropping the tildes) by rescaling $t\mapsto \lambda t$, $r\mapsto \lambda^{-1}r$, $y\mapsto \lambda^{-1}y$, $\psi\mapsto \lambda^3\psi$, $\zeta\mapsto\lambda^2\zeta$, $a\mapsto \lambda^{-2}a$, $\mu\mapsto \lambda^{-5}\mu$ and then taking the limit $\lambda\to 0$.

\end{enumerate}

\subsubsection{Subcase $b_{23}\neq0$: Myers-Perry metric with equal spins}

Here $p_0=2U^0b_{23}^2\neq0$ and $\epsilon=\mbox{sign}(U^0)$ (cf. \eqref{om_4_case2}, \eqref{constr_p0}, \eqref{def_eps}). Let us perform a transformation \eqref{rescaling}--\eqref{rescaling_param} accompanied by a further rescaling $(z,\bar z,y_2,\phi_2)\mapsto (\zeta,\bar\zeta,y,\chi)$ to dimensionless coordinates
	\be
	  y=\frac{y_2}{|b_{23}|} , \qquad \zeta=\lambda |b_{23}|z , \qquad \chi=\lambda^2 b_{23}\phi_2 , 
	\ee
with
\be
 \lambda=\sqrt{2|U^0|} .
\ee

Dropping the primes in \eqref{rescaling}--\eqref{rescaling_param}, the result for this case (cf. \eqref{hx45_case2}, \eqref{phi2_y2_case2}, \eqref{om_4_case2}, \eqref{hx23_case2}, \eqref{xi_23}, \eqref{P_case2}, \eqref{Q_case2_ii}, \eqref{xiu_4=0_case2_U0}, \eqref{xiu_23=0_case2}, \eqref{h0}) can be thus summarized as follows (up to rescaling $\bm_4\mapsto\mbox{sign}(b_{23})\bm_4$)
\beqn
 & & \omega^0_2=\omega^0_3=\omega^0_5=0 , \qquad \omega^0_4=b_{23}\sqrt{\epsilon(1-y^2)} , \\
 & & \xi^{u 0}_{2}=\xi^{u 0}_{3}=\xi^{u 0}_{5}=0 , \qquad \xi^{u 0}_{4}=\epsilon b_{23}\sqrt{\epsilon(1-y^2)} , \\
 & & \left(\xi^{\a 0}_{2}+i\xi^{\a 0}_{3}\right)\pa_\a=\frac{P}{\sqrt{\epsilon(1-y^2)}}\left(\pa_\zeta+\frac{\e i}{2\zeta P}\pa_\chi\right) , \label{CORRECTION} \\
 & & \xi^{\a 0}_{4}\pa_\a=\frac{1}{\sqrt{\epsilon(1-y^2)}}\pa_\chi , \qquad \xi^{\a 0}_{5}\pa_\a=\sqrt{\epsilon(1-y^2)}\pa_{y} , \\
 & & U^0=\frac{\epsilon}{2} , \qquad X^{A0}=\delta^A_u , 
\eeqn
where
\be
 \epsilon=\pm 1 , \qquad P=1+4\zeta\bar\zeta .
\ee 

After relabeling the only remaining integration constants as
\be
	b_{23}=a , \qquad \qquad \Phi_0=-6\mu , 
	\label{param_case2}
\ee
this shows that the present solution describes the {\em Myers-Perry metric with equal spins} of appendix~\ref{subsec_MP_equal} (after performing a spin in the plane $(23)$ to get rid of the extra phase $\e i\left(\zeta/\bar\zeta\right)^{1/2}$ in the first of \eqref{xi_23_MP_equal}).

\subsection{Case $\d b_{23}=0=\d b_{45}$: NUT metric with unequal NUTs}

Since both $b_{23}$ and $b_{45}$ are constant (with $b_{23}\neq b_{45}\neq 0$) one has $\omega^0_2=0=\omega^0_4$, so that, from \eqref{xi_b23}--\eqref{m_generic},  
\beqn
 & & \m{2}{4}{5}=\m{2}{5}{4}=\m{3}{4}{4}=\m{3}{5}{5}=0 , \label{m_NUT_1} \\
 & & \m{2}{5}{2}=\m{3}{5}{3}=\m{3}{4}{2}=\m{2}{4}{3}=0 , \label{m_NUT_2}
\eeqn
while \eqref{xi_om5} gives
\be
 U^0=0 .
 \label{U0=0_NUT}
\ee

The component $ijkl=2345$ of \eqref{11p_3} read
\be
 \xi^{\a 0}_{4}\m{2}{3}{5,\a}-\xi^{\a 0}_{5}\m{2}{3}{4,\a}=-\m{2}{3}{4}\m{4}{5}{4}-\m{2}{3}{5}\m{4}{5}{5} . 
 \label{11p_integrability_2}
\ee 

We can now use a $r$-and $u$-independent (23)-spin  \eqref{spin_23_m23k} (which does not affect \eqref{m_NUT_1}, \eqref{m_NUT_2}, \eqref{m_generic2}) to set 
\be
 \m{2}{3}{4}=0 , \qquad \m{2}{3}{5}=0 ,
	\label{spin_23_m23k_2}
\ee
since the integrability condition following from the two equations~\eqref{spin_23_m23k} is identically satisfied thanks to~\eqref{11p_integrability_2} with \eqref{xi_xi6}. Similarly, a (45)-spin  (which does not affect \eqref{spin_23_m23k_2}, \eqref{m_NUT_1}, \eqref{m_NUT_2}, \eqref{m_generic2}) can be used to set
\be
 \m{4}{5}{2}=0 , \qquad \m{4}{5}{3}=0 ,
	\label{spin_45_m45k_2}
\ee
thanks to the component $ijkl=4523$ of \eqref{11p_3} and \eqref{xi_xi1}.

The above frame choice produces a drastic simplification of eqs.~\eqref{xi_xi1}--\eqref{xi_xi6}, which thus read
\beqn
 & & 2\xi^{\b0}_{[2}\xi^{\a 0}_{3],\b}=-\xi^{\a 0}_{2}\m{2}{3}{2}-\xi^{\a 0}_{3}\m{2}{3}{3} , \qquad 2\xi^{\b0}_{[4}\xi^{\a 0}_{5],\b}=-\xi^{\a 0}_{4}\m{4}{5}{4}-\xi^{\a 0}_{5}\m{4}{5}{5} , \label{xi_xia_case3} \\
 & & 2\xi^{\b0}_{[2}\xi^{\a 0}_{4],\b}=0 , \qquad 2\xi^{\b0}_{[2}\xi^{\a 0}_{5],\b}=0 , \qquad 2\xi^{\b0}_{[3}\xi^{\a 0}_{4],\b}=0 , \qquad 2\xi^{\b0}_{[3}\xi^{\a 0}_{5],\b}=0 , \qquad  \label{xi_xib_case3}
\eeqn
while the remaining non-trivial components of \eqref{11p_3} reduce to
\beqn
 & & \xi^{\a 0}_{4}(\m{2}{3}{2}+i\m{2}{3}{3})_{,\a}=0 , \qquad \xi^{\a 0}_{5}(\m{2}{3}{2}+i\m{2}{3}{3})_{,\a}=0 , \qquad \xi^{\a0}_{3}\m{2}{3}{2,\a}-\xi^{\a0}_{2}\m{2}{3}{3,\a}=(\m{2}{3}{2})^2+(\m{2}{3}{3})^2 , \label{m232_y2_case3} \\
 & & \xi^{\a 0}_{2}(\m{4}{5}{4}+i\m{4}{5}{5})_{,\a}=0 , \qquad \xi^{\a 0}_{3}(\m{4}{5}{4}+i\m{4}{5}{5})_{,\a}=0 , \qquad \xi^{\a0}_{5}\m{4}{5}{4,\a}-\xi^{\a0}_{4}\m{4}{5}{5,\a}=(\m{4}{5}{4})^2+(\m{4}{5}{5})^2 .
\eeqn

Eqs.~\eqref{xi_xia_case3}, \eqref{xi_xib_case3} show that the distributions $\{\xi^{\a 0}_{2}\pa_\a,\xi^{\a 0}_{3}\pa_\a\}$ and $\{\xi^{\a 0}_{4}\pa_\a,\xi^{\a 0}_{5}\pa_\a\}$ are both integrable and that, furthermore, one can introduce two pairs of complex coordinates $(z,\bar z)$ and $(w,\bar w)$ such that $(\xi^{\a 0}_{2}+i\xi^{\a 0}_{3})\pa_\a=P(z,\bar z)\pa_z$ and $(\xi^{\a 0}_{4}+i\xi^{\a 0}_{5})\pa_\a=S(w,\bar w)\pa_w$ (up to a spin (23) independent of $(r,u,w,\bar w)$ and a spin (45) independent of $(r,u,z,\bar z)$, thus preserving \eqref{spin_23_m23k_2} and \eqref{spin_45_m45k_2}). 
The first of \eqref{xi_xia_case3} now gives $\m{2}{3}{2}+i\m{2}{3}{3}=-iP\bar P_{,z}/\bar P$, which, plugged into the last of \eqref{m232_y2_case3}, gives $P\bar P(\ln P\bar P)_{,z\bar z}=0$.
This means that the auxiliary 2-dimensional metric \eqref{2_metric} is flat. Since, by construction, the vectors $\xi^{\a 0}_{2}\pa_\a$ and $\xi^{\a 0}_{3}\pa_\a$ represent an orthonormal frame of \eqref{2_metric}, one can always replace the complex coordinates $(z,\bar z)$ by a real pair $(y_1,\phi_1)$ (more suitable for our purposes) such that (again up to a (23)-spin independent of $(r,u,w,\bar w)$),  $\xi^{\a 0}_{2}\pa_\a$ and $\xi^{\a 0}_{3}\pa_\a$ take the form
\be
  \xi^{\a 0}_{2}\pa_\a=\frac{1}{y_1}\pa_{\phi_1} , \qquad \xi^{\a 0}_{3}\pa_\a=\pa_{y_1} . \label{xi23_NUT}
\ee
Analogously, one can define $(y_2,\phi_2)$ such that, up to a (45)-spin (independent of $(r,u,y_1,\phi_1)$), 
\be
  \xi^{\a 0}_{4}\pa_\a=\frac{1}{y_2}\pa_{\phi_2} , \qquad \xi^{\a 0}_{5}\pa_\a=\pa_{y_2} . \label{xi45_NUT}
\ee

In this new frame, \eqref{xi_xia_case3} gives $\m{2}{3}{2}=-1/y_1$, $\m{2}{3}{3}=0$, $\m{4}{5}{4}=-1/y_2$, $\m{4}{5}{5}=0$. This will be used in the last step, which consists in determining the components $\xi^{u 0}_{i}$ using the first of~\eqref{xi_xi_2}.

Under a coordinate transformation 
\be
	u\mapsto u+V(y_1,\phi_1,y_2,\phi_2) ,
	\label{u_transf_NUT}
\ee
one has (with \eqref{xi23_NUT}, \eqref{xi45_NUT})
\be
  \xi^{u 0}_{2}\mapsto \xi^{u 0}_{2}+\frac{1}{y_1}V_{,\phi_1} , \qquad \xi^{u 0}_{3}\mapsto \xi^{u 0}_{3}+V_{,y_1} , \qquad \xi^{u 0}_{4}\mapsto \xi^{u 0}_{4}+\frac{1}{y_2}V_{,\phi_2} , \qquad \xi^{u 0}_{5}\mapsto \xi^{u 0}_{5}+V_{,y_2} . 
	\label{V2_NUT}
\ee
The component $ij=35$ of the first of~\eqref{xi_xi_2} gives $\xi^{u 0}_{5,y_1}-\xi^{u 0}_{3,y_2}=0$, which ensures that one can always find a function $V$ such as to set simultaneoulsy
\be
 \xi^{u 0}_{3}=0 , \qquad \xi^{u 0}_{5}=0 .
 \label{xiu_3_5_NUT}
\ee

Thanks to this, the remaining components of the first of~\eqref{xi_xi_2} reduce to
\beqn
 & & (y_1\xi^{u 0}_{2})_{,y_1}=-2b_{23}y_1 , \qquad (y_2\xi^{u 0}_{4})_{,y_2}=-2b_{45}y_2 , \\
 & & \xi^{u 0}_{2,y_2}=0, \qquad \xi^{u 0}_{4,y_1}=0, \qquad y_2\xi^{u 0}_{4,\phi_1}-y_1\xi^{u 0}_{2,\phi_2}=0 .
\eeqn 
These can be easily integrated to get $y_1\xi^{u 0}_{2}=-b_{23}y_1^2+g_1(\phi_1,\phi_2)$, $y_2\xi^{u 0}_{4}=-b_{45}y_2^2+g_2(\phi_1,\phi_2)$, with $g_{2,\phi_1}-g_{1,\phi_2}=0$. The latter condition ensures that we can use a transformation \eqref{u_transf_NUT}, \eqref{V2_NUT} with $V_{,y_1}=0=V_{,y_2}$ (thus preserving \eqref{xiu_3_5_NUT}) to set $g_1=0=g_2$, so that finally
\be
 \xi^{u 0}_{2}=-b_{23}y_1 , \qquad \xi^{u 0}_{4}=-b_{45}y_2 .
\label{xiu_2_4_NUT}
\ee

Summarizing, we have found that (cf. \eqref{xiu_3_5_NUT}, \eqref{xiu_2_4_NUT}, \eqref{xi23_NUT}, \eqref{xi45_NUT}, \eqref{U0=0_NUT})
\beqn
 & & \omega^0_i=0 , \qquad \xi^{u 0}_{2}=-b_{23}y_1 , \qquad \xi^{u 0}_{4}=-b_{45}y_2 \qquad \xi^{u 0}_{3}=0=\xi^{u 0}_{5} , \\
 & & \xi^{\a 0}_{2}\pa_\a=\frac{1}{y_1}\pa_{\phi_1} , \qquad \xi^{\a 0}_{3}\pa_\a=\pa_{y_1} , \qquad \xi^{\a 0}_{4}\pa_\a=\frac{1}{y_2}\pa_{\phi_2} , \qquad \xi^{\a 0}_{5}\pa_\a=\pa_{y_2} , \\ 
 & & U^0=0 , \qquad X^{A0}=\delta^A_u , 
\eeqn
which shows that this solution corresponds to the {\em NUT metric with two distinct NUT parameters} of appendix~\ref{app_case_e=0} (with $b_{23}=a_1$, $b_{45}=a_2$), first found in \cite{ManSte04} in different coordinates (cf. appendix~\ref{app_case_e=0} for details).

\section{Special case $b_{45}=b_{23}$ (shearfree solutions)}

\label{sec_shearfree}

\subsection{Subcase $b_{45}=b_{23}\neq0$ (shearfree, twisting solutions): NUT metric with equal NUTs}

\label{subsubsec_shearfree_twisting}

Let us define 
\be
 a\equiv b_{45}=b_{23}\neq0 .
\ee 
In this case \eqref{B_m_243} and \eqref{B_m_245} can be written compactly as
\be
  \m{3}{5}{k}=\m{2}{4}{k} , \qquad \m{3}{4}{k}=-\m{2}{5}{k} \qquad (k=2,3,4,5) , \label{m_b45=b23}
\ee
	so that \eqref{om2}--\eqref{b_der} give
	\be
	 \omega^0_2=\omega^0_3=\omega^0_4=\omega^0_5=0 , \qquad a=\mbox{const} .
	\ee
Using \eqref{xi_om_a} this implies
\be
 U^0=0 .
 \label{U0=0_b45=b23}
\ee	
	
The vectors $\xi^{\a 0}_i\pa_\a$ span a four-dimensional space and can be used to define an Euclidean (contravariant) metric $\xi^{\a 0}_i\xi^{\beta 0}_i\pa_\a\pa_\nb$ there (recall that $\xi^{\a 0}_{i,r}=0=\xi^{\a 0}_{i,u}$, cf.~\eqref{der_asym} and \eqref{om_xi_u}). The inverse of this auxiliary metric enables one to define the corresponding 1-forms $\eta_{\a0}^{i}\d x^\a$, such that
\be
 \xi^{\a 0}_i\eta_{\a0}^{j}=\delta^j_i .
 \label{eta_inverse}
\ee 

Using the second of~\eqref{omegai} (recall also  \eqref{l}, \eqref{n_adapted}, \eqref{L_shearfree} and the orthonormality relations of the frame vectors $\{\bl,\bn,\bm_i\}$), it is easy to see that $(m_a^{(2)}+im_a^{(3)})\d x^a=(r+ia)(\eta_{\a0}^{2}+i\eta_{\a0}^{3})\d x^\a$ and $(m_a^{(4)}+im_a^{(5)})\d x^a=(r+ia)(\eta_{\a0}^{4}+i\eta_{\a0}^{5})\d x^\a$. By considering the coefficient of the terms $\d x^\a\wedge\d x^\beta$ in the 2-form $\d\bm^i$ at the leading order in $r$ (with the definitions of the Ricci rotation coefficients $\M{i}{j}{k}$ of the full spacetime \eqref{Ricci_rot} and of their leading terms $\m{i}{j}{k}$, \eqref{L1i} or \eqref{ricci_asym1}), one concludes that 
\be
 \d\he^{i}_0=-\m{i}{j}{k}\he^{j}_0\wedge \he^{k}_0 .
 \label{1Cartan}
\ee	
This is the first Cartan equation in the auxiliary four-dimensional space, determining the connection 1-forms there in terms of the asymptotic rotation Ricci coefficients $\m{i}{j}{k}$. Thanks to this, one can use \eqref{11p_3} (with \eqref{U0=0_b45=b23}) to evaluate the Ricci identity in the auxiliary four-dimensional space, which implies that this space is flat.

The simplest choice is thus now to define adapted Cartesian coordinates such that (up to an $r$ and $u$-independent 4-rotation of the vectors $\xi^{\a 0}_i\pa_\a$)
\be
  \xi^{\a 0}_{i}\pa_\a=\pa_{x_i} .
\ee
This choice ensures that $\m{i}{j}{k}=0$, which simplifies the first of~\eqref{xi_xi_2}. It is easy to see that, using an appropriate transformation $u\mapsto u+V(x_2,x_3,x_4,x_5)$, the solution to the latter can always be written as
\be
  \xi^{u 0}_{2}=-2ax_3 , \qquad \xi^{u 0}_{4}=-2ax_5 , \qquad \xi^{u 0}_{3}=0=\xi^{u 0}_{5} .
\ee

Similarly as in section~\ref{subsubsec_MP_U0=0} (and \ref{subsubsec_b23=0}, case~\ref{b23=0_case2}), one can use the above asymptotic quantities to arrive at the following line-element
\be
 \d s^2=2\d r[\d u+2a(x_3\d x_2+x_5\d x_4)]+(r^2+a^2)(\d x_2^2+\d x_3^2+\d x_4^2+\d x_5^2)+\frac{\mu r}{(r^2+a^2)^2}[\d u+2a(x_3\d x_2+x_5\d x_4)]^2 .
\ee

For $\mu\neq 0$, a transformation~\eqref{du_NUT} shows that this metric is equivalent to the {\em doubly NUT metric with equal NUT parameters} (49) of \cite{ManSte04} (setting their $\lambda=0$ and $n_1=n_2$), in turn equivalent to our line-element~\eqref{MP_e=0_stationary} with $a_1=a_2$ (see appendix~\ref{app_case_e=0} for more details).

\subsection{Subcase $b_{45}=b_{23}=0$ (shearfree, twistfree solutions): Schwarzschild-Tangherlini metric}

\label{subsec_RT}

When $b_{45}=b_{23}=0$, the geodesic null congruence defined by $\bl$ is expanding, shearfree and twistfree (cf.~\eqref{inverse} with \eqref{b_eigenf}), i.e., one has a Robinson-Trautman spacetime. This case has been excluded from the analysis of the previous sections, where we assumed $b_{[ij]}\neq0$. However, it was already fully explored in \cite{PodOrt06}. Thanks to the assumption~\eqref{assump_C2}, the corresponding line-element can thus be written as \cite{PodOrt06}
\be
	\d s^2=-\left(2U^0-\frac{\mu}{r^3}\right)\d u^2+2\d u\d r+r^2h_{\a\nb}(x)\d x^\a\d x^\nb ,
	\label{RT}
\ee
where $h_{\a\nb}(x)$ is the metric of a four-dimensional space of constant curvature having a Ricci scalar ${\cal R}=24U^0$, and $2U^0=0,\pm1$ (see also \eqref{MP_6D_equal} with $a=0$ and \eqref{MP_e=0}  with $a_1=0=a_2$). The case $2U^0=1$ corresponds to the static, spherically symmetric Schwarzschild-Tangherlini black hole \cite{Tangherlini63}.

For completeness, let us mention that, in this case, the extra assumption~\eqref{assump_C2} can be dropped in the following sense (see section~8 of \cite{OrtPraPra13} and section~5 of \cite{ReaGraTur13}): all vacuum spacetimes admitting a {\em twistfree, non-degenerate mWAND} in six dimensions belong to the Robinson-Trautman family (and are thus of type D). Such a result holds also in the presence of a cosmological constant \cite{OrtPraPra13,ReaGraTur13} (see \cite{PodOrt06} for the corresponding line-element) and also in five dimensions \cite{Ortaggioetal12}, but not for $n>6$ (counterexamples being known \cite{OrtPraPra13}).

\section*{Acknowledgments}

I am grateful to Vojt\v ech Pravda and Alena Pravdov\' a for collaboration at an early stage of this work. I also thank Gabriel Bernardi de Freitas for comments on \cite{deFGodRea15}. This work has been supported by research plan RVO: 67985840 and research grant GA\v CR 13-10042S. The stay of the author at Instituto de Ciencias F\'{\i}sicas y Matem\'aticas, Universidad Austral de Chile has been supported by CONICYT PAI ATRACCI{\'O}N DE CAPITALE HUMANO AVANZADO DEL EXTRANJERO Folio 80150028.

\renewcommand{\thesection}{\Alph{section}}
\setcounter{section}{0}

\renewcommand{\theequation}{{\thesection}\arabic{equation}}

\section{Myers-Perry metrics in six dimensions}
\setcounter{equation}{0}

\label{app_MP}

Let us consider the following six-dimensional metric in the ``Eddington-like'' coordinates $(u,r,\mu_1,\mu_2,\phi_1,\phi_2)$ 
\beqn
 \d s^2=-\e\d u^2+2\d r(\d u+\e a_1\mu_1^2\d\phi_1+\e a_2\mu_2^2\d\phi_2)+(r^2+a_1^2)(\d\mu_1^2+\mu_1^2\d\phi_1^2)+(r^2+a_2^2)(\d\mu_2^2+\mu_2^2\d\phi_2^2)+\e r^2\d\mu_3^2  \nonumber \\
	{}+\frac{\mu r}{\rho^2}(\d u+\e a_1\mu_1^2\d\phi_1+\e a_2\mu_2^2\d\phi_2)^2 ,
	\label{MP_6D}
\eeqn
where $a_1$, $a_2$ and $\mu$ are constants,  and 
\beqn
 & & \rho^2=\e\mu_1^2r^2(r^2+a_2^2)+\e\mu_2^2r^2(r^2+a_1^2)+\mu_3^2(r^2+a_1^2)(r^2+a_2^2) , \label{P} \\
 & & \mu_3^2=1-\e(\mu_1^2+\mu_2^2) , \qquad \e=\pm 1 . \label{constr_MP}
\eeqn

The six-dimensional Myers-Perry metric corresponds to the case $\e=+1$ (cf. eq.~(3.20) of \cite{MyePer86}, after defining $u=t+r$). From this, the metric with $\e=-1$ (still a vacuum solution, not considered in \cite{MyePer86}) can be obtained by the complex coordinate transformation $u\mapsto iu$, $r\mapsto -ir$, $\mu_1\mapsto i\mu_1$, $\mu_2\mapsto i\mu_2$ along with the rescaling $a_1\mapsto ia_1$, $a_2\mapsto ia_2$, $\mu\mapsto -i\mu$ (a similar transformation works in any even dimension). For $r\to\infty$, the parameter $\e$ specifies the norm of the Killing vector field $\pa_u$, as well as the sign of the curvature of 4-spaces of constant $r$ and $u$ (cf. \cite{Kinnersley69} and section~IV of \cite{KinnersleyPhD} in four dimensions).

The line-element~\eqref{MP_6D} is manifestly of the Kerr-Schild form, and the Kerr-Schild geodesic null vector $\ell_a\d x^a=\d u+\e a_1\mu_1^2\d\phi_1+\e a_2\mu_2^2\d\phi_2$ is a multiple WAND \cite{OrtPraPra09}. One can check that the corresponding contravariant vector is simply (cf. \cite{MyePer86})
\be
	\bl=\pa_r , 
	\label{l_MP}
\ee	
affinely parametrized by the coordinate $r$.

\subsection{Generic case: two unequal spins ($a_1^2-a_2^2\neq 0\neq a_1a_2$)} 

\label{app_MP_generic}

In this case it is useful to replace the coordinates $(\mu_1,\mu_2)$ with $(y_1,y_2)$ via (similarly as in \cite{CheLuPop06}) 
\be
 \mu_1^2=\e\frac{(a_1^2-y_1^2)(a_1^2-y_2^2)}{a_1^2(a_1^2-a_2^2)} , \qquad \mu_2^2=\e\frac{(a_2^2-y_1^2)(a_2^2-y_2^2)}{a_2^2(a_2^2-a_1^2)} .
 \label{jacobi}
\ee
This transformation diagonalizes the ``$\mu_i$-part'' of the generalized Myers-Perry metric \eqref{MP_6D}, which thus becomes 
\beqn
 \d s^2=-\e\d u^2+2\d r(\d u+\e a_1\mu_1^2\d\phi_1+\e a_2\mu_2^2\d\phi_2)+(r^2+y_1^2)g_1\d y_1^2+(r^2+y_2^2)g_2\d y_2^2+(r^2+a_1^2)\mu_1^2\d\phi_1^2+(r^2+a_2^2)\mu_2^2\d\phi_2^2  \nonumber \\
	{}+\frac{\mu r}{\rho^2}(\d u+\e a_1\mu_1^2\d\phi_1+\e a_2\mu_2^2\d\phi_2)^2 ,
	\label{MP_6D_CLP}
\eeqn
with
\be
 g_1=\frac{\e(y_2^2-y_1^2)}{(a_1^2-y_1^2)(a_2^2-y_1^2)} , \qquad g_2=\frac{\e(y_1^2-y_2^2)}{(a_1^2-y_2^2)(a_2^2-y_2^2)} , \qquad \rho^2=(r^2+y_1^2)(r^2+y_2^2) . 
 \label{g1_g2}
\ee
This spacetime is known to be of Weyl type D \cite{Hamamotoetal07,PraPraOrt07} and {\em not} purely electric (i.e., $\Phi^A_{ij}\neq 0$) \cite{OrtPraPra09,HerOrtWyl13}. It follows \cite{Wylleman12} (see also Remark~3.15 of \cite{HerOrtWyl13}) that it admits precisely two distinct multiple WANDs (of course, one of these is parallel to $\bl$), which possess the same optical properties (thanks to the reflection symmetry argument of \cite{PraPraOrt07}, related to a comment in footnote~3 of \cite{Kinnersley69}). It can also be observed that if one rescales $\phi_1\mapsto a_1\phi_1$, $\phi_2\mapsto a_2\phi_2$, then for $\e=-1$ one can replace $a_1^2$ (or $a_2^2$, but not both, since $g_1$ and $g_2$ in \eqref{g1_g2} must be both positive for a Lorentzian signature) in the above metric by an arbitrary parameter which can have any sign, or vanish, still giving rise to a vacuum solution.

Together with \eqref{l_MP}, a frame parallely transported along $\bl$ is given by
\beqn
 & & \bn=\pa_u+\frac{\e\rho^2-\mu r}{2\rho^2}\pa_r , \\
 & & \bm_2=\frac{1}{\sqrt{g_1}}\frac{r}{r^2+y_1^2}\left(\e\pa_u+\pa_r-\frac{y_1}{r}\pa_{y_1}-\frac{\e a_1}{a_1^2-y_1^2}\pa_{\phi_1}-\frac{\e a_2}{a_2^2-y_1^2}\pa_{\phi_2}\right) , \\
 & & \bm_3=\frac{1}{\sqrt{g_1}}\frac{y_1}{r^2+y_1^2}\left(\e\pa_u+\pa_r+\frac{r}{y_1}\pa_{y_1}-\frac{\e a_1}{a_1^2-y_1^2}\pa_{\phi_1}-\frac{\e a_2}{a_2^2-y_1^2}\pa_{\phi_2}\right) , \\
 & & \bm_4=\frac{1}{\sqrt{g_2}}\frac{r}{r^2+y_2^2}\left(\e\pa_u+\pa_r-\frac{y_2}{r}\pa_{y_2}-\frac{\e a_1}{a_1^2-y_2^2}\pa_{\phi_1}-\frac{\e a_2}{a_2^2-y_2^2}\pa_{\phi_2}\right) , \\
 & & \bm_5=\frac{1}{\sqrt{g_2}}\frac{y_2}{r^2+y_2^2}\left(\e\pa_u+\pa_r+\frac{r}{y_2}\pa_{y_2}-\frac{\e a_1}{a_1^2-y_2^2}\pa_{\phi_1}-\frac{\e a_2}{a_2^2-y_2^2}\pa_{\phi_2}\right) .
\eeqn

The asymptotic quantities (cf. \eqref{der_asym} with $n=6$) are thus given by
\beqn
 & & \omega^0_2=\frac{1}{\sqrt{g_1}} , \qquad \omega^0_4=\frac{1}{\sqrt{g_2}} , \qquad \omega^0_3=0=\omega^0_5 , \\
 & & \xi^{u 0}_{2}=\e\omega^0_2 , \qquad \xi^{u 0}_{4}=\e\omega^0_4 , \qquad \xi^{u 0}_{3}=0=\xi^{u 0}_{5} , \\
 & & \xi^{\a 0}_{2}\pa_\a=-\e\omega^0_2\left(\frac{a_1}{a_1^2-y_1^2}\pa_{\phi_1}+\frac{a_2}{a_2^2-y_1^2}\pa_{\phi_2}\right) , \qquad \xi^{\a 0}_{3}\pa_\a=\omega^0_2\pa_{y_1} , \\
 & & \xi^{\a 0}_{4}\pa_\a=-\e\omega^0_4\left(\frac{a_1}{a_1^2-y_2^2}\pa_{\phi_1}+\frac{a_2}{a_2^2-y_2^2}\pa_{\phi_2}\right) , \qquad \xi^{\a 0}_{5}\pa_\a=\omega^0_4\pa_{y_2} , \\
 & & U^0=\frac{\e}{2} , \qquad \Phi_0=-6\mu , \qquad X^{A0}=\delta^A_u .
\eeqn

One can also verify that $b_{23}=y_1$, $b_{45}=y_2$.

A Boyer-Lindquist form of metric~\eqref{MP_6D_CLP} can be obtained by defining  new coordinates $(t,\psi_1,\psi_2)$
\be
 \d u=\d t+\frac{(r^2+a_1^2)(r^2+a_2^2)}{\e(r^2+a_1^2)(r^2+a_2^2)-\mu r}\d r , \quad \d\phi_1=\d\psi_1-\frac{a_1(r^2+a_2^2)}{\e(r^2+a_1^2)(r^2+a_2^2)-\mu r}\d r , \quad \d\phi_2=\d\psi_2-\frac{a_2(r^2+a_1^2)}{\e(r^2+a_1^2)(r^2+a_2^2)-\mu r}\d r ,
\ee 
so that 
\beqn
 \d s^2=-\e\d t^2+\frac{\rho^2}{\e(r^2+a_1^2)(r^2+a_2^2)-\mu r}\d r^2+(r^2+y_1^2)\e\frac{y_2^2-y_1^2}{(a_1^2-y_1^2)(a_2^2-y_1^2)}\d y_1^2+(r^2+y_2^2)\e\frac{y_1^2-y_2^2}{(a_1^2-y_2^2)(a_2^2-y_2^2)}\d y_2^2 \nonumber \\
{}+(r^2+a_1^2)\e\frac{(a_1^2-y_1^2)(a_1^2-y_2^2)}{a_1^2(a_1^2-a_2^2)}\d\psi_1^2+(r^2+a_2^2)\e\frac{(a_2^2-y_2^2)(a_2^2-y_1^2)}{a_2^2(a_2^2-a_1^2)}\d\psi_2^2  \nonumber \\
{}+\frac{\mu r}{\rho^2}\left[\d t+\frac{(a_1^2-y_1^2)(a_1^2-y_2^2)}{a_1(a_1^2-a_2^2)}\d\psi_1+\frac{(a_2^2-y_1^2)(a_2^2-y_2^2)}{a_2(a_2^2-a_1^2)}\d\psi_2\right]^2 .
	\label{MP_6D_CLP_stationary}
\eeqn

For $\e=+1$, metric~\eqref{MP_6D_CLP_stationary} coincides with metric (45) of \cite{CheLuPop06} (apart from a trivial rescaling of $\psi_1$ and $\psi_2$, and after setting $g=L_1=L_2=0$ in \cite{CheLuPop06}; a transformation to the alternative metric form~(48) of \cite{CheLuPop06}, which allows also for $\e=-1$, requires instead linear redefinitions of the Killing coordinates similar to (47) of \cite{CheLuPop06}). In these cooordinates, 
\be
	\ell_{a}^\pm\d x^a=\d t\pm\frac{\rho^2}{\e(r^2+a_1^2)(r^2+a_2^2)-\mu r}\d r+\frac{(a_1^2-y_1^2)(a_1^2-y_2^2)}{a_1(a_1^2-a_2^2)}\d\psi_1+\frac{(a_2^2-y_1^2)(a_2^2-y_2^2)}{a_2(a_2^2-a_1^2)}\d\psi_2 ,
\ee
define the two multiple WANDs.

\subsection{Two equal spins ($a_1^2=a_2^2$)} 

\label{subsec_MP_equal}

Clearly one cannot use \eqref{jacobi} when $a_1=a_2\equiv a$. Let us now define, instead, the dimensionless coordinates $(y_1,y_2)$ via 
\be
 \mu_1^2=\e(1-y_2^2)y_1^2 , \qquad  \mu_2^2=\e(1-y_2^2)(1-y_1^2) ,
 \label{jacobi_}
\ee
such that $\mu_3^2=y_2^2$. Metric \eqref{MP_6D} becomes 
\beqn
 \d s^2=-\e\d u^2+2\d r\left[\d u+\e a(\mu_1^2\d\phi_1+\mu_2^2\d\phi_2)\right]+(r^2+a^2y_2^2)\frac{\e\d y_2^2}{1-y_2^2}+(r^2+a^2)\left[\e(1-y_2^2)\frac{\d y_1^2}{1-y_1^2}+\mu_1^2\d\phi_1^2+\mu_2^2\d\phi_2^2\right]  \nonumber \\
	{}+\frac{\mu r}{\rho^2}\left[\d u+\e a(\mu_1^2\d\phi_1+\mu_2^2\d\phi_2)\right]^2 ,
	\label{MP_6D_equal}
\eeqn
with
\be
 \rho^2=(r^2+a^2)(r^2+a^2y_2^2) .
\ee
(Let us note that metric~\eqref{MP_6D_equal} can also be obtained from \eqref{MP_6D_CLP} by replacing $y_1\mapsto\sqrt{a_1^2-(a_1^2-a_2^2)y_1^2}$, $y_2\mapsto a_1y_2$, and then taking the limit $a_2\to a_1\equiv a$.)\footnote{Slightly different coordinates for metric~\eqref{MP_6D_equal} with $\e=+1$ were employed in appendix~C of \cite{Gibbonsetal05} (after setting $\lambda=0$ there). The only non-obvious transformation required to map these two coordinate systems into each other involves a redefinition of $\phi_1$ and $\phi_2$ as in (3.22) of \cite{MyePer86} with $\mu=0$.}

A frame parallely transported along $\bl$ is given by \eqref{l_MP} with 
\beqn
 & & \bn=\pa_u+\frac{\e\rho^2-\mu r}{2\rho^2}\pa_r , \\
 & & \bm_2=\frac{r}{r^2+a^2}\sqrt{\frac{1-y_1^2}{\e(1-y_2^2)}}\left(-\e\frac{a}{r}\pa_{y_1}-\frac{y_1}{1-y_1^2}\pa_{\phi_1}+\frac{1}{y_1}\pa_{\phi_2}\right) , \\
 & & \bm_3=\frac{r}{r^2+a^2}\sqrt{\frac{1-y_1^2}{\e(1-y_2^2)}}\left[\e \pa_{y_1}+\frac{a}{r}\left(-\frac{y_1}{1-y_1^2}\pa_{\phi_1}+\frac{1}{y_1}\pa_{\phi_2}\right)\right] , \label{m3_equal} \\
 & & \bm_4=\frac{ra\sqrt{\e(1-y_2^2)}}{r^2+a^2y_2^2}\left[\e\pa_u+\pa_r-\frac{y_2}{r}\pa_{y_2}-\frac{\e}{a(1-y_2^2)}\left(\pa_{\phi_1}+\pa_{\phi_2}\right)\right] , \\
 & & \bm_5=\frac{a^2y_2\sqrt{\e(1-y_2^2)}}{r^2+a^2y_2^2}\left[\e\pa_u+\pa_r+\frac{r}{a^2y_2}\pa_{y_2}-\frac{\e}{a(1-y_2^2)}(\pa_{\phi_1}+\pa_{\phi_2})\right] . 
\eeqn

From this and \eqref{der_asym} one can read off the following asymptotic quantities
\beqn
 & & \omega^0_2=\omega^0_3=\omega^0_5=0 , \qquad \omega^0_4=a\sqrt{\e(1-y_2^2)} , \label{om4_equal} \\
 & & \xi^{u 0}_{2}=\xi^{u 0}_{3}=\xi^{u 0}_{5}=0 , \qquad \xi^{u 0}_{4}=\e a\sqrt{\e(1-y_2^2)} , \\
 & & \xi^{\a 0}_{2}\pa_\a=\frac{1}{\sqrt{\e(1-y_2^2)}}\left(-\frac{y_1}{\sqrt{1-y_1^2}}\pa_{\phi_1}+\frac{\sqrt{1-y_1^2}}{y_1}\pa_{\phi_2}\right) , \qquad \xi^{\a 0}_{3}\pa_\a=\e\sqrt{\frac{1-y_1^2}{\e(1-y_2^2)}}\pa_{y_1} , \\
 & & \xi^{\a 0}_{4}\pa_\a=-\frac{1}{\sqrt{\e(1-y_2^2)}}\left(\pa_{\phi_1}+\pa_{\phi_2}\right) , \qquad \xi^{\a 0}_{5}\pa_\a=\sqrt{\e(1-y_2^2)}\pa_{y_2} , \\
 & & U^0=\frac{\e}{2} , \qquad \Phi_0=-6\mu , \qquad X^{A0}=\delta^A_u .
\eeqn

It is useful to observe that the complex coordinate transformation $(y_1,\phi_1,\phi_2)\mapsto(\zeta,\bar\zeta,\chi)$ defined by
\be
	\zeta=-\frac{1}{2}e^{\e i(\phi_1-\phi_2)}\frac{\sqrt{1-y_1^2}}{y_1} , \qquad \chi=-\phi_1 ,
\ee
allows one to rewrite (the remaining quantities being unchanged)
\be
 \left(\xi^{\a 0}_{2}+i\xi^{\a 0}_{3}\right)\pa_\a=\frac{\e i\left(\zeta/\bar\zeta\right)^{1/2}P}{\sqrt{\e(1-y_2^2)}}\left(\pa_\zeta+\frac{\e i}{2\zeta P}\pa_\chi\right) , \qquad \xi^{\a 0}_{4}\pa_\a=\frac{1}{\sqrt{\e(1-y_2^2)}}\pa_\chi , 
\label{xi_23_MP_equal}
\ee
where
\be
  P\equiv\frac{1}{y_1^2}=1+4\zeta\bar\zeta .
\ee 

Here one has $b_{23}=a$, $b_{45}=ay_2$. For $a=0$, the twist and shear of $\bl$ vanish and \eqref{MP_6D_equal} becomes the 6D Schwarzschild-Tangherlini metric \cite{Tangherlini63} with a transverse space of positive ($\epsilon=+1$) or negative ($\e=-1$) constant curvature, written in Robinson-Trautman coordinates \cite{PodOrt06}.

Boyer-Lindquist coordinates may be defined by 
\be
 \d u=\d t+\frac{(r^2+a^2)^2}{\e(r^2+a^2)^2-\mu r}\d r , \qquad \d\phi_1=\d\psi_1-\frac{a(r^2+a^2)}{\e(r^2+a^2)^2-\mu r}\d r , \qquad \d\phi_2=\d\psi_2-\frac{a(r^2+a^2)}{\e(r^2+a^2)^2-\mu r}\d r ,
\ee 
which gives
\beqn
 & & \d s^2=-\e\d t^2+\frac{\rho^2}{\e(r^2+a^2)^2-\mu r}\d r^2+\frac{\mu r}{\rho^2}\left[\d t+a(1-y_2^2)y_1^2\d\psi_1+a(1-y_2^2)(1-y_1^2)\d\psi_2\right]^2  \nonumber \\
 & & \qquad\qquad {}+(r^2+a^2y_2^2)\e\frac{\d y_2^2}{1-y_2^2}+(r^2+a^2)\e(1-y_2^2)\left[\frac{\d y_1^2}{1-y_1^2}+y_1^2\d\psi_1^2+(1-y_1^2)\d\psi_2^2\right] .
 \label{BL_equal}
\eeqn
This metric form is not contained in (48) of \cite{CheLuPop06}, since there both twist eigenvalues $b_{23}$ and $b_{45}$ are used as coordinates, which is not possible here (but see the comments above for the limit from \eqref{MP_6D_CLP} to \eqref{MP_6D_equal}, and for the relation between the Boyer-Lindquist form of the former, i.e., \eqref{MP_6D_CLP_stationary}, and (48) of \cite{CheLuPop06}). This spacetime is of Weyl type D and, in the cooordinates \eqref{BL_equal}, the unique \cite{Wylleman12,HerOrtWyl13} two multiple WANDs are given by (cf.~\cite{PraPraOrt07,OrtPraPra09})
\be
	\ell_{a}^\pm\d x^a=\d t\pm\frac{\rho^2}{\e(r^2+a^2)^2-\mu r}\d r+a(1-y_2^2)y_1^2\d\psi_1+a(1-y_2^2)(1-y_1^2)\d\psi_2 .
\ee

\subsection{Single spin ($a_1\neq0$, $a_2=0$)} 

\label{subsec_MP_single}

When $a_2=0$, let us relabel $a_1\equiv a$ and replace the coordinates $(\mu_1,\mu_2,\phi_1,\phi_2)$ with $(y,\chi,\zeta,\bar\zeta)$ via 
\be
	\phi_1=-a\chi , \qquad e^{-2\e i\phi_2}=\zeta/\bar\zeta , \qquad\mu_1^2=\e\frac{a^2-y^2}{a^2} , \qquad  \mu_2^2=y^2\frac{4\zeta\bar\zeta}{P^2}, \qquad P\equiv 1+\e a^2\zeta\bar\zeta . 
 \label{jacobi_single}
\ee
Metric \eqref{MP_6D} becomes 
\beqn
 \d s^2=-\e\d u^2+2\d r\left(\d u-\e a^2\mu_1^2\d\chi\right)+(r^2+y^2)\frac{\e\d y^2}{a^2-y^2}+(r^2+a^2)a^2\mu_1^2\d\chi^2+r^2y^2\frac{4\d\zeta\d\bar\zeta}{P^2}+\frac{\mu r}{\rho^2}\left(\d u-\e a^2\mu_1^2\d\chi\right)^2 ,
	\label{MP_6D_single}
\eeqn
with
\be
 \rho^2=r^2(r^2+y^2) .
\ee
(Metric~\eqref{MP_6D_single} can also be obtained from \eqref{MP_6D_CLP} by replacing $y_2\mapsto a_2y_2$, taking the limit $a_2\to 0$, and then defining $\zeta=a_1^{-1}e^{-i\e\phi_2}\sqrt{\e(1-y_2)(1+y_2)^{-1}}$, $\chi=-a_1^{-1}\phi_1$ -- similar limits have been recently considered in \cite{Krtousetal16}). We observe that, in these coordinates, the parameter $a$ can also be purely imaginary when $\e<0$, i.e., values $a^2\le0$ are permitted. For $\e=+1$, \eqref{MP_6D_single} equals metric~(3.1) of \cite{MyePer86} after defining $y=a\cos\theta$ and trivially rescaling $\chi$.

A frame parallely transported along $\bl$ is given by
\beqn
 & & \bn=\pa_u+\frac{\e\rho^2-\mu r}{2\rho^2}\pa_r , \\
 & & \bm_2+i\bm_3=\frac{1}{r}\frac{P}{y}\pa_\zeta , \\
 & & \bm_4=\frac{r}{r^2+y^2}\sqrt{\e(a^2-y^2)}\left(\e\pa_u+\pa_r-\frac{y}{r}\pa_{y}+\frac{\e}{a^2-y^2}\pa_{\chi}\right) , \\
 & & \bm_5=\frac{y}{r^2+y^2}\sqrt{\e(a^2-y^2)}\left(\e\pa_u+\pa_r+\frac{r}{y}\pa_{y}+\frac{\e}{a^2-y^2}\pa_{\chi}\right) ,
\eeqn
from which one extracts the asymptotic quantities (cf. \eqref{der_asym}) 
\beqn
 & & \omega^0_2=\omega^0_3=\omega^0_5=0 , \qquad \omega^0_4=\sqrt{\e(a^2-y^2)} ,  \\
 & & \xi^{u 0}_{2}=\xi^{u 0}_{3}=\xi^{u 0}_{5}=0 , \qquad \xi^{u 0}_{4}=\e\sqrt{\e(a^2-y^2)} ,  \\
 & & \left(\xi^{\a 0}_{2}+i\xi^{\a 0}_{3}\right)\pa_\a=\frac{P}{y}\pa_\zeta , \\
 & & \xi^{\a 0}_{4}\pa_\a=\frac{1}{\sqrt{\e(a^2-y^2)}}\pa_\chi , \qquad \xi^{\a 0}_{5}\pa_\a=\sqrt{\e(a^2-y^2)}\pa_{y} , \\
 & & U^0=\frac{\e}{2} , \qquad \Phi_0=-6\mu , \qquad X^{A0}=\delta^A_u .
\eeqn

One finds $b_{23}=0$, $b_{45}=y$.

Defining new coordinates $(t,\psi)$ \cite{MyePer86}
\be
 \d u=\d t+\frac{r^2+a^2}{\e(r^2+a^2)-\frac{\mu}{r}}\d r , \qquad \d \chi=\d\psi+\frac{1}{\e(r^2+a^2)-\frac{\mu}{r}}\d r ,
\ee
one obtains from~\eqref{MP_6D_single} a Boyer-Lindquist form of the metric 
\beqn
 \d s^2=-\e\d t^2+(r^2+y^2)\left[\frac{\d r^2}{\e(r^2+a^2) -\frac{\mu}{r}}+\frac{\e\d y^2}{a^2-y^2}\right]+(r^2+a^2)\e (a^2-y^2)\d\psi^2+r^2y^2\frac{4\d\zeta\d\bar\zeta}{P^2}+\frac{\mu r}{\rho^2}\left[\d t-(a^2-y^2)\d\psi\right]^2 ,
	\label{MP_6D_single_stat}
\eeqn

This is cf.~(3.5a) of \cite{MyePer86} (with $y=a\cos\theta$) for $\e=+1$, and (2.5) of \cite{Klemm98} (with $y=a\cosh\theta$) for $\e=-1$ (after rescaling $\psi\mapsto\psi/a$, specializing \cite{MyePer86,Klemm98} to six dimensions, and setting the cosmological constant to zero in \cite{Klemm98}). In both cases, at $r=0$ there is a curvature singularity (where the Kretschmann scalar diverges). 
This spacetime is of Weyl type D and, in the cooordinates \eqref{MP_6D_single_stat}, the unique \cite{Wylleman12,HerOrtWyl13} two multiple WANDs are given by (cf.~\cite{PraPraOrt07,OrtPraPra09})
\be
	\ell_{a}^\pm\d x^a=\d t\pm\frac{r^2+y^2}{\e(r^2+a^2)-\frac{\mu}{r}}\d r-(a^2-y^2)\d\psi .
\ee

Another useful parametrization is obtained by defining a new time coordinate $\tilde t$
\be
 \d t=\d\tilde t+a^2\d\psi ,
\ee
which gives
\beqn
 \d s^2=\frac{1}{r^2+y^2}\left[-\Delta_r(\d\tilde t+y^2\d\psi)^2+\e(a^2-y^2)(\d\tilde  t-r^2\d\psi)^2\right]
+(r^2+y^2)\left[\frac{\d r^2}{\Delta_r}+\frac{\e\d y^2}{a^2-y^2}\right]+r^2y^2\frac{4\d\zeta\d\bar\zeta}{P^2} ,
	\label{MP_6D_single_stat_2}
\eeqn
with
\be
 \Delta_r=\e(r^2+a^2)-\frac{\mu}{r} .
\ee
For $\e=+1$, \eqref{MP_6D_single_stat_2} corresponds to the line-element (5.16), (5.17) of \cite{Krtousetal16} (in the case of vanishing cosmological constant and NUT parameter).

\subsection{Non-spinning metric ($a_1=0=a_2$)} 

For $a_1=a_2=0$, metric \eqref{MP_6D} reduces to the Schwarzschild-Tangherlini solution \cite{Tangherlini63} in Robinson-Trautman coordinates \cite{PodOrt06} ($\d\mu_1^2+\mu_1^2\d\phi_1^2+\d\mu_2^2+\mu_2^2\d\phi_2^2+\e\d\mu_3^2$, with \eqref{constr_MP}, being the metric of a 4-space of constant curvature with sign $\e=\pm 1, 0$). This is the only case when $\bl$ is {\em twistfree}. For $\e=\pm 1$, different forms of the metric can be obtained by setting $a=0$ in \eqref{MP_6D_equal} or \eqref{BL_equal}, while, for $\e=0$, by setting $a_1=0=a_2$ in the metric \eqref{MP_e=0} or \eqref{MP_e=0_stationary} given below. These spacetimes are also of Weyl type D.

\subsection{The NUT limit and metrics with a shearfree $\bl$} 

\label{app_case_e=0}

A different vacuum solution can be obtained as a limit of \eqref{MP_6D}. Performing the coordinate and parameters rescaling 
	\beqn
	  & & r=\lambda^{-1}r' , \qquad u=\lambda u' , \qquad \mu_1=\lambda y_1 , \qquad \mu_2=\lambda y_2 , \\ 
		& & a_1=\lambda^{-1} a_1' , \qquad a_2=\lambda^{-1} a_2' , \qquad \mu=\lambda^{-5}\mu' , 
	\eeqn 
and then setting $\lambda=0$, one arrives at the line-element (dropping the primes)\footnote{After the limit, the sign of $\e$ has no meaning (just redefine $\phi_1\mapsto\e\phi_1$ and $\phi_2\mapsto\e\phi_2$) so we can set $\e=1$. Alternatively, one can arrive at \eqref{MP_e=0} by substituting $r=\lambda^{-1}r'$, $u=\lambda u'$, $y_1^2=a_1^2 (\lambda^{-2}-y_1'^2)$, $y_2^2=a_2^2 (\lambda^{-2}-y_2'^2)$, 
		$a_1=\lambda^{-1}a_1'$, $a_2'=\lambda^{-1} a_2'$, $\mu=\lambda^{-5}\mu'$ in \eqref{MP_6D_CLP} and then setting $\lambda=0$.}	
\beqn
 \d s^2=2\d r(\d u+a_1y_1^2\d\phi_1+a_2y_2^2\d\phi_2)+(r^2+a_1^2)(\d y_1^2+y_1^2\d\phi_1^2)+(r^2+a_2^2)(\d y_2^2+y_2^2\d\phi_2^2) \nonumber \\
	{}+\frac{\mu r}{\rho^2}(\d u+a_1y_1^2\d\phi_1+a_2y_2^2\d\phi_2)^2 ,
	\label{MP_e=0}
\eeqn
with
\be
 \rho^2=(r^2+a_1^2)(r^2+a_2^2) .
\ee

A frame parallely transported along $\bl$ is given by \eqref{l_MP} with
\beqn
 & & \bn=\pa_u-\frac{\mu r}{2\rho^2}\pa_r , \\
 & & \bm_2=\frac{r}{r^2+a_1^2}\left(-a_1y_1\pa_u-\frac{a_1}{r}\pa_{y_1}+\frac{1}{y_1}\pa_{\phi_1}\right) , \\
 & & \bm_3=\frac{a_1}{r^2+a_1^2}\left(-a_1y_1\pa_u+\frac{r}{a_1}\pa_{y_1}+\frac{1}{y_1}\pa_{\phi_1}\right) , \\
 & & \bm_4=\frac{r}{r^2+a_2^2}\left(-a_2y_2\pa_u-\frac{a_2}{r}\pa_{y_2}+\frac{1}{y_2}\pa_{\phi_2}\right) , \\
 & & \bm_5=\frac{a_2}{r^2+a_2^2}\left(-a_2y_2\pa_u+\frac{r}{a_2}\pa_{y_2}+\frac{1}{y_2}\pa_{\phi_2}\right) .
\eeqn

This gives the asymptotic quantities 
\beqn
 & & \omega^0_2=\omega^0_3=\omega^0_4=\omega^0_5=0 , \\
 & & \xi^{u 0}_{2}=-a_1y_1 , \qquad \xi^{u 0}_{4}=-a_2y_2 , \qquad \xi^{u 0}_{3}=0=\xi^{u 0}_{5} , \\
 & & \xi^{\a 0}_{2}\pa_\a=\frac{1}{y_1}\pa_{\phi_1} , \qquad \xi^{\a 0}_{3}\pa_\a=\pa_{y_1} , \qquad \xi^{\a 0}_{4}\pa_\a=\frac{1}{y_2}\pa_{\phi_2} , \qquad  \xi^{\a 0}_{5}\pa_\a=\pa_{y_2} , \\
 & & U^0=0 , \qquad \Phi_0=-6\mu , \qquad X^{A0}=\delta^A_u .
\eeqn

Note that here $b_{23}=a_1$, $b_{45}=a_2$. Therefore {\em $\bl$ is shearfree iff $a_2=a_1$}. It is, additionally, twistfree iff $a_2=a_1=0$, in which case \eqref{MP_e=0} becomes the 6D Schwarzschild-Tangherlini metric with a flat transverse space \cite{Tangherlini63,PodOrt06}. 

It may be useful to give the line-element \eqref{MP_e=0} also in Boyer-Lindquist coordinates. Defining $t$ via
\be
 \d u=\d t-\frac{\rho^2}{\mu r}\d r ,
\label{du_NUT}
\ee
one obtains
\beqn
 \d s^2=\frac{\mu r}{\rho^2}(\d t+a_1y_1^2\d\phi_1+a_2y_2^2\d\phi_2)^2-\frac{\rho^2}{\mu r}\d r^2+(r^2+a_1^2)(\d y_1^2+y_1^2\d\phi_1^2)+(r^2+a_2^2)(\d y_2^2+y_2^2\d\phi_2^2) ,
	\label{MP_e=0_stationary}
\eeqn
so that the Killing vector field $\pa_t$ is timelike for $\mu r<0$. The apparent singularity at $r=0$ (not present in the coordinates \eqref{MP_e=0}) is a Killing horizon of $\pa_t$ (with flat spatial sections). (For $\mu=0$ these coordinates are singular, but in that case \eqref{MP_e=0} is just Minkowski spacetime and $\bn$ becomes covariantly constant.) Metric~\eqref{MP_e=0_stationary} is equivalent to the NUT solution (49) of \cite{ManSte04} if the cosmological constant is set to zero there (after defining Cartesian coordinates in each of the two 2-planes $(y_1,\phi_1)$, $(y_2,\phi_2)$ and after a simple shift of $t$). A 4D analog of \eqref{MP_e=0_stationary} is given by the NUT metric with a flat transverse space \cite{NewTamUnt63} (cf. also section~12.3.2 of \cite{GriPodbook}).

In these coordinates, $\ell_a\d x^a=\d t-\frac{\rho^2}{\mu r}\d r+a_1y_1^2\d\phi_1+a_2y_2^2\d\phi_2$ and $n_a\d x^a=\frac{\mu r}{2\rho^2}(\d t+\frac{\rho^2}{\mu r}\d r+a_1y_1^2\d\phi_1+a_2y_2^2\d\phi_2)$. 
Since \eqref{MP_e=0_stationary} is invariant under $t\mapsto-t$, $\phi_1\mapsto-\phi_1$, $\phi_2\mapsto-\phi_2$, it follows \cite{PraPraOrt07} that $\bn$ points along a second multiple WAND, with the same optical properties of $\bl$. In particular, the Weyl tensor is thus of type D and not purely electric \cite{OrtPraPra09,HerOrtWyl13}. It follows \cite{Wylleman12,HerOrtWyl13} that $\bl$ and $\bn$ define the unique pair of multiple WANDs. Different type D vacuum solutions admitting a pair of shearfree mWANDs are contained in \cite{AwaCha02,ManSte04}, as discussed in \cite{OrtPraPra13} -- they do not belong to the class studied in the present paper because they violate the assumption~\eqref{assump_C2}.

\section{Newman-Penrose formalism when $\bl$ is a non-degenerate mWAND}
\setcounter{equation}{0}

\label{app_NP}

Here we summarize the notation and the equations of the Newman-Penrose formalism needed in the present paper. The Ricci rotation coefficients $L_{ab}$, $N_{ab}$ and $\M{i}{a}{b}$ are defined by \cite{Pravdaetal04}
\be
 L_{ab}=\ell_{a;b} , \qquad N_{ab}=n_{a;b}  , \qquad \M{i}{a}{b}=m^{(i)}_{a;b} ,
 \label{Ricci_rot}
\ee
and satisfy the identities $L_{0a}=N_{1a}=N_{0a}+L_{1a}=\M{i}{0}{a} + L_{ia} = \M{i}{1}{a}+N_{ia}=\M{i}{j}{a}+\M{j}{i}{a}=0$. As in the main text, $\bl$ is a geodesic and affinely parametrized mWAND, and we use a frame parallelly transported along $\bl$, so that the following quantities vanish identically
\be
 L_{i0}=0 , \qquad L_{10}=0, \qquad \M{i}{j}{0}=0, \qquad N_{i0}=0 . 
\label{parall_transp}
\ee
Moreover, thanks to \eqref{detL}, we can set (cf. \eqref{Li1}) 
\be
  L_{i1}=0 . 
\ee

Covariant derivatives along the frame vectors are denoted as
\be
	D \equiv \ell^a \nabla_a, \qquad \T\equiv n^a \nabla_a, \qquad \delta_i \equiv m_{(i)}^{a} \nabla_a . 
 \label{covder}
\ee
 
For the non-vanishing frame components of the Weyl tensor (i.e., those of non-positive b.w.) we define the symbols \cite{Pravdaetal04,PraPraOrt07}
\beqn
 & & \WD{ij} = C_{0i1j} , \qquad \WDS{ij} =\WD{(ij)} , \qquad \WDA{ij} =\WD{[ij]} , \qquad \WD{ } =\WD{ii} , \label{def_Phi} \\
 & & \Psi_{i} = C_{101i}, \qquad \Psi_{ijk}= \frac{1}{2} C_{1kij}, \qquad \Psi_{ij} = \frac{1}{2} C_{1i1j} , \label{def_Psi} 
\eeqn 
which satisfy the identities $C_{01ij}=2 C_{0[i|1|j]}=2\WDA{ij} $, $2C_{0(i|1|j)}=2\WDS{ij} =-C_{ikjk}$, $2C_{0101}= -C_{ijij}=2\WD{ } $, $\Psi_i=2 \Psi_{ijj}$, $\Psi_{\{ijk\}}=0$, $\Psi_{ijk}=-\Psi_{jik}$, $\Psi_{ij}=\Psi_{ji}$, and $\Psi_{ii}=0$. Throughout the paper, the vacuum Einstein equations $R_{ab}=0$ hold, so that $R_{abcd}=C_{abcd}$.

Under the above conditions, the Bianchi equations (B.1), (B.6), (B.9) and (B.4) of \cite{Pravdaetal04} (cf. also (16)-(18), (22) of \cite{OrtPraPra09b}) take the simplified form
\BEA
	& & D\Psi_i=-2\Psi_s L_{si} +\delta_i \WD{} \label{PII-B1}, \\
	& &  2D\Psi_{ijk}= -2\Psi_{ijs}L_{sk}-\Psi_{i}L_{jk}+\Psi_{j}L_{ik}-2\delta_k \WDA{ij} -4\WDA{[i|s} \Ms_{|j]k}\label{PII-B6}, \\
	& & D\Psi_{jki}=2\Psi_{[k|si}L_{s|j]}+\Psi_{i}L_{[jk]}-\delta_{[k}\WD{j]i} +\WD{[k|s} \Ms_{i|j]}-\WD{si} \Ms_{[jk]}, \label{PII-B9} \\
	& & 2D\Psi_{ij}=-2\Psi_{is}L_{sj} +\delta_{j}\Psi_i+\Psi_i L_{1j}+\Psi_s \Ms_{ij} \nonumber \\ 
	& & \qquad\qquad							{}+\Delta \WD{ji} +\WD{} N_{ij}-2\WDA{is} N_{sj}+\WD{si} N_{sj}+\WD{js} \Ms_{i1}+\WD{si} \Ms_{j1} , \label{PII-B4}
\EEA
while (B.13,\cite{Pravdaetal04}) becomes
\beqn
  & & -\T C_{ijkm}+4\delta_{[k|}\Psi_{ij|m]} = 2\Psi_{im} L_{jk} +  4\Psi_{[j|k} L_{|i]m} -2\Psi_{jm} L_{ik}+4\Psi_{[i|sk} \Ms_{|j]m}+  4\Psi_{ij[k|} L_{1|m]}  \nonumber \\ 
  & & \qquad\qquad  {}+  4\Psi_{[j|sm} \Ms_{|i]k} +  4\Psi_{ijs} \Ms_{[km]} +  2C_{ij[k|s} \Ms_{|m]1} +  2C_{[i|skm} \Ms_{|j]1}  \nonumber \\ 
	& & \qquad\qquad {}- 4\WDA{ij} N_{[km]}   +  2\Phi_{[i|m} N_{|j]k}  +  2\Phi_{[j|k} N_{|i]m}+  2C_{ij[k|s} N_{s|m]} .
\label{B13}
\eeqn 

The commutators \cite{Coleyetal04vsi} read
\beqn
 & & \Delta D-D\Delta=L_{11}D , \label{comm_DelD} \\
 & & \delta_i D-D\delta_i=L_{1i}D+L_{ji}\delta_j , \label{comm_dD} \\
 & & \delta_i\Delta-\Delta\delta_i=N_{i1}D-L_{1i}\Delta+(N_{ji}+\M{j}{i}{1})\delta_j , \label{comm_dDel} \\
 & & \delta_{[i}\delta_{j]}=N_{[ij]}D+L_{[ij]}\Delta+\M{k}{[i}{j]}\delta_k . \label{comm_dd}
\eeqn

The Ricci identities (11k), (11i), (11h), (11l),  (11o) and (11p) of \cite{OrtPraPra07} take the form\footnote{A correction to (11p,\cite{OrtPraPra07}), pointed out in footnote~7 of \cite{OrtPraPra13rev}, has no consequences in the present paper.} 
\beqn
	& &     \delta_{[j|} L_{i|k]} =  L_{1[j|} L_{i|k]} +  L_{il} \M{l}{[j}{k]} + L_{l[j|}  \M{l}{i|}{k]}  , \label{11k_NP} \\
	& &   \T L_{ij} =  L_{11} L_{ij}-L_{kj}\M{k}{i}{1}- L_{ik} (N_{kj} + \M{k}{j}{1}) - \Phi_{ij}  ,  \label{11i_NP} \\
  & & \T N_{ij} - \delta_j N_{i1} = - L_{11} N_{ij}+2N_{i1}L_{1j} + N_{k1}\M{k}{i}{j}-N_{kj}\M{k}{i}{1} - N_{ik} (N_{kj} + \M{k}{j}{1}) - 2\Psi_{ij} ,  \label{11h_NP} \\
	& & \delta_{[j|} N_{i|k]} =  - L_{1[j|} N_{i|k]} + N_{i1} L_{[jk]}   +  N_{il} \M{l}{[j}{k]} + N_{l[j|}  \M{l}{i|}{k]}- \Psi_{jki}  , \label{11l_NP} \\
	& &  \T \M{i}{j}{k} - \delta_{k} \M{i}{j}{1} =  N_{j1} L_{ik}- L_{jk} N_{i1} + \M{i}{j}{1}L_{1k} + \M{i}{l}{1} \M{l}{j}{k}-\M{i}{l}{k} \M{l}{j}{1} - \M{i}{j}{l} (N_{lk} + \M{l}{k}{1}) - 2\Psi_{ijk} ,  \label{11o_NP} \\
& &  \delta_{[k|} \M{i}{j|}{l]} 
=  N_{i[l|} L_{j|k]}  
+ L_{i[l|} N_{j|k]} +L_{[kl]} \M{i}{j}{1} +   \M{i}{p}{[k|} \M{p}{j|}{l]} + \M{i}{j}{p}  \M{p}{[k}{l]} - \textstyle{\frac{1}{2}} C_{ijkl}  . \label{11p_NP}
\eeqn

%

\end{document}